\documentclass[sigplan, 10pt, nonacm]{acmart}

\AtBeginDocument{%
  }

\usepackage{tikz}
\usepackage{amsmath}

\usepackage{filecontents}
\usepackage{xspace}
\usepackage{threeparttable}
\usepackage{makecell}
\usepackage{url}
\usepackage{cleveref}

\usepackage{tikz}
\usepackage{threeparttable}
\usepackage{soul}
\usepackage{colortbl}
\usepackage{multirow}
\usepackage{tabularx}
\usepackage{amsmath,pifont, mathtools}
\usepackage{amsmath}
\usepackage[overload]{empheq}
\usepackage{nccmath}
\usepackage{cases}
\usepackage{graphicx}
\usepackage{pifont}
\usepackage{wasysym}
\usepackage{mathtools}
\usepackage{float}
\usepackage{subcaption}
\usepackage{adjustbox}
\usepackage{makecell}%
\usepackage{tikz}
\usepackage{booktabs}
\usepackage{cleveref}
\usepackage{comment}
\usepackage[normalem]{ulem}
\usepackage{enumitem}
\usepackage{listings}
\usepackage{color}
\usepackage{textcomp}
\usepackage{balance}

\usepackage{algorithm}
\usepackage{algpseudocode}

\usepackage{booktabs}
\usepackage{caption}
\usepackage{tikz}
\usepackage{xspace}
\usepackage{pifont}
\newcommand{\xmark}{\ding{55}}
\usepackage{url}
\usepackage[font=small]{caption}
\definecolor{red}{rgb}{1,0,0}
\definecolor{gray}{rgb}{0.5,0.5,0.5}

\crefformat{section}{#2§#1#3}
\Crefformat{section}{#2§#1#3}
\crefrangeformat{section}{#3§#1#4--#5§#2#6}
\crefmultiformat{section}{#2§#1#3}{ and #2§#1#3}{, #2§#1#3}{, and #2§#1#3}
\crefformat{subsection}{#2§#1#3}
\Crefformat{subsection}{#2§#1#3}
\crefrangeformat{subsection}{#3§#1#4--#5§#2#6}
\crefmultiformat{subsection}{#2§#1#3}{ and #2§#1#3}{, #2§#1#3}{, and #2§#1#3}
\crefformat{appendix}{#2§#1#3}
\Crefformat{appendix}{#2§#1#3}
\crefrangeformat{appendix}{#3§#1#4--#5§#2#6}
\crefmultiformat{appendix}{#2§#1#3}{ and #2§#1#3}{, #2§#1#3}{, and #2§#1#3}
\crefformat{subappendix}{#2§#1#3}
\Crefformat{subappendix}{#2§#1#3}
\crefrangeformat{subappendix}{#3§#1#4--#5§#2#6}
\crefmultiformat{subappendix}{#2§#1#3}{ and #2§#1#3}{, #2§#1#3}{, and #2§#1#3}

\newcommand\hmm[1]{\ifnum\ifhmode\spacefactor\else2000\fi>1000 \uppercase{#1}\else#1\fi}
\newcommand{\sys}{Frontier\xspace}

\begin{document}


\title{\sys: Towards Comprehensive and Accurate \\LLM Inference Simulation}
% \title{\sys: Simulating the Next Generation of LLM Inference Systems}
% \hx{need a better title}

\author{
\vspace{1em}
{\rm Yicheng Feng$^{\text{1}}$}\quad
{\rm Xin Tan$^{\text{1}}$}\quad
{\rm Yangtao Deng$^{\text{1}}$}\quad
{\rm Yimin Jiang$^{\text{2}}$}\quad
{\rm Yibo Zhu$^{\text{3}}$}\quad
{\rm Hong Xu$^{\text{1}}$}\protect\\[0.95em]
\textit{$^{\text{1}}$The Chinese University of Hong Kong}\qquad
\textit{$^{\text{2}}$Anuttacon}\qquad
\textit{$^{\text{3}}$StepFun}
\vspace{3em}
}

%!TEX root = newmain.tex

\begin{abstract}

Modern LLM serving is no longer homogeneous or monolithic. Production systems now combine disaggregated execution, complex parallelism, runtime optimizations, and stateful workloads such as reasoning, agents, and RL rollouts. Simulation is attractive for exploring this growing design space, yet existing simulators lack the architectural completeness and decision-grade fidelity it demands. Their monolithic-replica abstractions are ill-suited to disaggregated serving, while average-case analytical proxies can distort SLA predictions and even reverse optimization conclusions.

We present \sys, a discrete-event simulator for modern LLM inference serving. \sys features a disaggregated abstraction. It captures the structure and dynamics of modern serving systems by modeling co-location, Prefill-Decode Disaggregation (PDD), and Attention-FFN Disaggregation (AFD) with role-specific cluster workers, incorporating key runtime optimizations (e.g., CUDA Graphs, speculative decoding) within the scheduler-batch-engine loop, and supporting stateful requests for emerging workloads. It further provides accurate and generalizable predictions of computation, communication, and memory costs across diverse serving scenarios with complex workload compositions. On 16-H800 GPU testbed, \sys achieves an average throughput error below 4\%. Compared with state-of-the-art simulators, it reduces end-to-end latency error from 44.9\% to 6.4\% under co-location and from 51.7\% to 2.6\% under disaggregation. It scales to over 1K GPUs on commodity CPUs and enables new use cases such as SLA-dependent Pareto frontier exploration, heterogeneous disaggregated allocation, agentic reasoning scheduling validation, and RL post-training reconfiguration.
We release Frontier at \url{https://github.com/NetX-lab/Frontier}.

\end{abstract}

\maketitle
\thispagestyle{plain}
\pagestyle{plain}

% \input{abstract}
%!TEX root = newmain.tex

\section{Introduction}
\label{sec:introduction}

Inference has become a first-order production cost for LLM service providers amid the explosive growth of token use from not just chatbots, but also coding and computer use agents, which is particularly evident since 2026~\cite{agrawal2026revatitransparentgpufreetimewarp,xu2026aiconfiguratorlightningfastconfigurationoptimization}. 
% Large services
% now generate tokens continuously, and their latency, capacity, and cloud
% budget all depend on how efficiently those tokens are served~\cite{agrawal2026revatitransparentgpufreetimewarp,xu2026aiconfiguratorlightningfastconfigurationoptimization}.
This pressure has driven the evolution of serving systems in multiple fronts: The \textit{serving architecture} has moved beyond the traditional co-located paradigm, where both the prefill and decode phases execute within a single monolithic engine, to the disaggregated architectures.
Most notably, Prefill-Decode Disaggregation (PDD) decouples the compute-bound prefill from the memory-bound decode and places them on separate GPU pools
connected via KV-cache transfers~\cite{zhong2024distserve,dynamo}.
Pushing this further, Attention-FFN Disaggregation (AFD) separates the attention
and feed-forward computations during the decode phase, routing activations between distinct hardware pools~\cite{stepfun2025step3largeaffordablemodelsystem,zhu2025megascale}.
Alongside these architectural shifts, deployments are heavily augmented with complex \textit{parallelism strategies} (tensor (TP)~\cite{shoeybi2019megatron}, pipeline (PP)~\cite{narayanan2019pipedream}, data (DP), and expert (EP)~\cite{lepikhin2020gshard} parallelism) and diverse \textit{runtime optimizations}---continuous batching~\cite{agrawal2024taming}, PagedAttention~\cite{kwon2023efficient}, CUDA Graphs~\cite{CUDAGraph}, prefix caching~\cite{zheng2024sglang}, chunked prefill~\cite{agrawal2023sarathi}, and speculative decoding~\cite{leviathan2023fastinferencetransformersspeculative}.
% Together, these architectural and runtime mechanisms fundamentally alter batch shapes, memory footprints,
% and the underlying scheduler states that govern request admission and batching..

The serving workloads are also becoming more diverse at the same time.
Sparse MoE~\cite{li2023accelerating,singh2023hybrid,liu2024deepseek} models are widely used that activates a different subset of experts for each token at each layer. 
Reasoning models produce long hidden reasoning chains before yielding the final answer~\cite{guo2025deepseek}.
Agentic systems introduce complex tool calls and multi-turn session states~\cite{yao2022react,wang2024survey}.
Furthermore, RL rollouts for post-training create workload bursts, long decode tails, and shifting phase mixes~\cite{tan2026orchestrrl,zhang2026heddle}.
% To navigate this complexity, service providers must tune parallelism strategies, cluster ratios, hardware GPU types, batching policies, memory budgets, and runtime features under strict Service Level Agreements (SLAs)---particularly regarding Time-To-First-Token (TTFT) and Time-Per-Output-Token (TPOT).
% This immense configuration space can easily exceed $10{,}000$ permutations~\cite{xu2026aiconfiguratorlightningfastconfigurationoptimization}.
% As a result, even modest real-cluster parameter sweeps consume thousands of GPU-hours and can cost tens to hundreds of thousands of dollars~\cite{agrawal2026revatitransparentgpufreetimewarp}.
Consequently, the interplay of complex workloads, advanced runtime optimizations, and disaggregated serving architectures has made production deployments exceedingly difficult to model, reason, and optimize.

As a result, simulation as an essential performance modeling approach has received increasing attention~\cite{agrawal2024vidur,agrawal2026revatitransparentgpufreetimewarp,xu2026aiconfiguratorlightningfastconfigurationoptimization,cho2026llmservingsim}.
It provides a cheap and fast way to characterize a serving system without the need for expensive real deployment (at scale)~\cite{agrawal2024vidur,agrawal2026revatitransparentgpufreetimewarp,xu2026aiconfiguratorlightningfastconfigurationoptimization}.
% It can narrow the hardware search
% ex-situ, test scheduling policies before rollout, and validate whether a
% candidate deployment can meet TTFT and TPOT targets.
% Discrete-event simulators are the common tool for this purpose: they re-implement the
% system's control logic, predict the duration of each GPU batch, advance
% virtual time, and process the next event~\cite{agrawal2024vidur,agrawal2026revatitransparentgpufreetimewarp}.
% \hx{to do, depending on the rest of the flow}
Existing simulators, however, fall short in two dimensions that we believe make them fundamentally ineffective as a performance modeling tool in many practically crucial scenarios. 
% \hx{this claim needs to be explicit and strong but also precise}

% \textit{First, most simulators lack modeling completeness.}
\textit{First, architectural completeness.}
Most simulators (e.g., Vidur~\cite{agrawal2024vidur}, APEX~\cite{lin2024apex}) lack support for disaggregated serving and many of the runtime optimizations discussed above, and these capabilities are not easy to retrofit.
The root cause is architectural: they assume a set of homogeneous, monolithic replicas executing uniform operations.
That abstraction is fundamentally mismatched to disaggregated systems, where different GPU pools execute distinct roles in parallel and interact through explicit data transfer and synchronization.
% For instance, failing to model CUDA Graph execution inherently introduces a substantial prediction error,
% as this mechanism alone dictates a $37.1\%\mathrm{-}59.7\%$ reduction in TPOT for disaggregated serving setups.
% For example, introducing disaggregated architectures into LLM serving can allow the system to serve up to $7.4\times$ more requests compared to a co-located serving system~\cite{zhong2024distserve},
% while CUDA Graphs can further reduce TPOT by up to $37.1\%$ (\cref{subsec:challenges}).
% However, existing simulators cannot be easily expanded to support these techniques.
% The key problem is that most of these works (e.g., Vidur~\cite{agrawal2024vidur}, APEX~\cite{}) design their systems around a homogeneous-replica assumption:
% they model serving systems as collections of isolated, monolithic engine replicas executing uniform operations.
% This abstraction mismatch makes them fundamentally incompatible with disaggregated architectures, where different GPU pools execute distinct workloads with specific behaviors in parallel and interact with each other through explicit transfer and synchronization edges.
% and simple analytical models fail to capture their complexity.
Likewise, runtime optimizations are entangled with scheduler-visible state and therefore must be modeled as first-class behavior rather than analytical patches.
Approaches such as AIConfigurator~\cite{xu2026aiconfiguratorlightningfastconfigurationoptimization} can therefore incur large errors and even predict the wrong optimization trend (\cref{subsec:challenges}).
More broadly, because serving workloads are stateful and phase-changing~\cite{yao2022react,wang2024survey}, treating inference as an open-loop sequence of regular operator executions is architecturally incomplete.

\textit{Second, decision-grade fidelity.}
Existing simulators usually deliver coarse average-case accuracy, but production use demands predictions that remain faithful under realistic serving dynamics~\cite{agrawal2024vidur}.
As discussed, LLM serving workloads are dynamic and stateful: continuous batching, variable-length requests, KV-cache growth, prefill/decode asymmetry, scheduling interference, and memory pressure jointly shape latency, memory usage, and queue evolution.
In this closed loop, even modest per-operator or per-batch errors can become decision-changing: they can shift time-to-first-token (TTFT)/time-per-output-token (TPOT) or memory estimates across service level agreement (SLA) boundaries, distort the Pareto frontier, and lead to the wrong configuration choice (\cref{subsec:challenges}).
The core problem is that existing inference simulators smooth these dynamics into average-case models, especially for operator runtime and KV-cache budgeting: token-count proxies can cause $32.6\%$ errors on FlashAttention and $21.0\%$ on MoE GroupedGEMM, while ``total minus weights'' memory models overestimate KV-cache budget by up to $27.2\%$ and throughput by up to $32.4\%$ (\cref{subsec:challenges}).
\begin{comment}
We build \sys, a discrete-event simulator (DES) for inference that addresses these limitations.
Architecturally, \sys uses a control plane that compiles a serving specification
into role-specific cluster workers for co-location, PDD, and AFD.
It explicitly expresses the domain layout of disaggregated systems and models the dependencies introduced by each role split,
including pipeline stages, PDD KV-cache transfer, AFD activation transfer, and MoE EP synchronization.
\sys provides feature-specific adapters inside the scheduler--batch-engine loop.
These adapters update the scheduler-visible state for prefix caching, the batch shape and timing path
for CUDA Graph, the per-request token progress for speculative decoding, and the prompt progress for chunked prefill.
\sys also introduces a stateful request abstraction for new workloads such as reasoning,
tool-using agents, and RL rollouts; each request can carry thinking rounds, tool-call delays, per-round token plans, and phase transitions.
For fidelity,
\sys introduces a fidelity plane that resolves operator, collective, transfer,
and KV-cache budget queries using calibrated operator predictors and profiled KV-cache budget models.
This design lets \sys represent new architectures and runtime features
without hard-coding a new simulator for each case.
\end{comment}
% ============================================================================
% Revised version (top-down, concise, maps to the two shortcomings)
% ============================================================================

We build \sys,\footnote{We release Frontier at \url{https://github.com/NetX-lab/Frontier}.} a discrete-event simulator that addresses both limitations through two complementary design principles:
(1) for \textit{architectural completeness}, \sys abandons the monolithic-replica abstraction.
A control plane compiles a serving specification into \textit{role-specific cluster workers} that explicitly model the domain layout and synchronization dependencies of co-located, PDD, and AFD architectures.
Runtime optimizations are captured as \textit{feature-specific adapters} inside the scheduler--batch-engine loop, directly updating scheduler-visible state rather than applied as analytical afterthoughts.
To represent increasingly diverse workloads, \sys introduces a \textit{stateful request abstraction} where each request can carry thinking rounds, tool-call delays, per-round token plans, and phase transitions, naturally modeling reasoning, agentic, and RL-rollout workloads.  
(2) for \textit{decision-grade fidelity}, \sys introduces a \textit{fidelity plane} that replaces coarse average-case proxies with calibrated, hardware-aware predictors.
Operator runtimes, collective costs, transfer delays, and KV-cache budgets are each resolved through profiled models grounded in actual CUDA kernel behavior and memory footprints, closing the error gaps that make average-case approximations decision-changing.
% ============================================================================

We implemented \sys with roughly 70K LoC in Python. 
On a 16-GPU H800 testbed, we evaluate \sys across dense and MoE models, serving architectures, and varied workloads.
\sys achieves average throughput errors below 4\% in both co-located and disaggregated settings with modern runtime optimizations enabled.
On the SharedGPT trace, \sys reduces the average end-to-end (E2E) error from 44.9\% to 6.4\% under co-location, and from 51.7\% to 2.6\% under PDD/AFD, compared with state-of-the-art simulators.

Beyond fidelity, \sys scales to large-scale clusters (e.g., 1K+ GPUs) even on commodity CPU machines, making it practical to study deployment questions that prior simulators cannot reproduce as configured.
Under traditional non-reasoning batch serving, \sys enables SLA-aware Pareto-frontier analysis across co-location, PDD, and AFD, and evaluates heterogeneous PDD/AFD allocations by separating nominal cost reduction from SLA-safe efficiency.
Under emerging stateful workloads, \sys supports mechanism studies for multi-round agentic reasoning schedulers and quantifies rollout-tail reconfiguration opportunities for RL post-training (\cref{sec:use_cases}).

%!TEX root = newmain.tex
\section{Background and Motivation}
\label{sec:motivation}

\begin{table}[t]
\centering
\small
\definecolor{mygreen}{rgb}{0.1, 0.6, 0.1}
\definecolor{mutedred}{rgb}{0.72, 0.36, 0.36}
\newcommand{\cmark}{\textcolor{mygreen}{\checkmark}}
\newcommand{\rmark}{\textcolor{mutedred}{\xmark}}
\setlength{\tabcolsep}{4.5pt}
\renewcommand{\arraystretch}{0.95}
\begin{tabular}{@{}l ccccc@{}}
\toprule
 & \textbf{\sys} & AC & LS & VD & AP \\
\midrule
\rowcolor[gray]{0.9} \multicolumn{6}{@{}l}{\textit{\textbf{Architecture}}} \\
Co-located Serving & \cmark & \cmark & \cmark & \cmark & \cmark \\
PDD Serving & \cmark & \cmark & \cmark & \rmark & \rmark \\
AFD Serving & \cmark & \rmark & \rmark & \rmark & \rmark \\
\midrule
\rowcolor[gray]{0.9} \multicolumn{6}{@{}l}{\textit{\textbf{Parallelism}}} \\
PP / TP / DP & \cmark & \cmark & \cmark & \cmark & \cmark \\
MoE / EP & \cmark & \cmark & \cmark & \rmark & -- \\
DP Attention & \cmark & -- & \rmark & \rmark & \rmark \\
\midrule
\rowcolor[gray]{0.90} \multicolumn{6}{@{}l}{\textit{\textbf{Runtime Optimizations}}} \\
Speculative Decoding / MTP & \cmark & -- & \rmark & \rmark & \rmark \\
CUDA Graph & \cmark & \rmark & \rmark & \rmark & \rmark \\
Prefix Caching & \cmark & -- & \cmark & \rmark & \rmark \\
Quantization & \cmark & \cmark & \rmark & \rmark & \cmark \\
Chunked Prefill & \cmark & -- & \rmark & \cmark & \rmark \\
Hierarchical Caching & \cmark & \rmark & \cmark & \rmark & \rmark \\
\midrule
\rowcolor[gray]{0.9} \multicolumn{6}{@{}l}{\textit{\textbf{Simulation Features}}} \\
Op/Mem. Fidelity & \cmark & -- & -- & -- & -- \\
Ex-situ Simulation & \cmark & \cmark & \cmark & \rmark & \cmark \\
Thinking/Reasoning Modeling & \cmark & \rmark & \rmark & \rmark & \rmark \\

\bottomrule
\end{tabular}
\caption{Comparison with representative simulators. AC: AIConfigurator~\cite{xu2026aiconfiguratorlightningfastconfigurationoptimization}, LS: LLMServingSim2.0~\cite{cho2026llmservingsim}, VD: Vidur~\cite{agrawal2024vidur}, AP: APEX~\cite{lin2024apex}. \cmark: supported, \rmark: unsupported, --: partial/limited.}
\label{tab:work_compare}
\vspace{-0.5em}
\end{table}

\subsection{Why Inference Simulation Matters}
\label{subsec:why}

% \noindent The economics of LLM serving have shifted dramatically. 
% Global token generation and consumption have grown by orders of 
% magnitude in the past two years~\cite{agrawal2026revatitransparentgpufreetimewarp,xu2026aiconfiguratorlightningfastconfigurationoptimization}, and inference now dominates 
% the total cost of deploying a production model. This pressure 

The LLM serving has pushed industry toward aggressive system--model--hardware co-design~\cite{stepfun2025step3largeaffordablemodelsystem,guo2025deepseek}.
The design space is exploding.
For a 200B-parameter MoE model, plausible configurations 
easily reach 1k+ combinations~\cite{xu2026aiconfiguratorlightningfastconfigurationoptimization}.
On a 64-GPU H100 cluster at current cloud pricing, 
sweeping 100 configurations takes 12{,}800--25{,}600 GPU-hours and 
costs up to \$180{,}000~\cite{agrawal2026revatitransparentgpufreetimewarp,agrawal2024vidur}. 
% Worse, a configuration tuned to one 
% workload often becomes brittle when request patterns, model versions, 
% or hardware shift---forcing operators to repeat the sweep. 
In practice, most teams abandon systematic tuning and fall back on 
conservative defaults, leaving $3$--$5\times$ throughput on the 
table~\cite{xu2026aiconfiguratorlightningfastconfigurationoptimization,miao2022galvatron,zheng2022alpa}. 

A well-built simulator provides value far beyond replacing expensive hardware sweeps. 
By making hardware--software interactions explicit and event-level state observable, it can support Pareto search under SLA constraints, 
policy prototyping and diagnosis, dynamic workload analysis, and pre-deployment validation of heterogeneous placement or disaggregation plans. 
In practice, simulators can help one team map the frontier, another debug an agentic scheduler or RL rollout policy, 
and a third check whether a disaggregated deployment still behaves as intended before rollout.

\subsection{Challenges}
\label{subsec:challenges}
Table~\ref{tab:work_compare} summarizes the comparison between \sys and existing simulators.\footnote{In cases of discrepancy between the paper's claims and the latest open-source repository as of April. 2026, we refer to the latter.}
Building upon the analysis of existing simulators in \cref{sec:introduction}, we identify two primary challenges that must be addressed to realize a practical simulation framework.
% We frame these as two coupled challenges.
% Despite active progress, existing simulators~\cite{agrawal2024vidur,xu2026aiconfiguratorlightningfastconfigurationoptimization,cho2026llmservingsim,lin2024apex}, exhibit two fundamental shortcomings for modern LLM serving. 
% First, the majority cover only a narrow subset of optimizations, lacking the capability to simulate practical serving architectures. 
% Second, even when such support exists (e.g., AIConfigurator), it typically relies on analytical modeling to approximate advanced techniques. 
% This approach comes at the expense of both fidelity and expressiveness, exhibiting prediction errors of up to $10\times$ (\cref{subsec:e2e_fidelity}).
% We frame these as two coupled challenges.
Unless specified, experiments use vLLM v0.10.2 (V1 engine) on an 8$\times$A800 GPU server.
% The first concerns \textit{what happens}: whether the simulator generates the correct serving events and state transitions. 
% The second concerns \textit{what it costs}: whether each event carries the right latency and memory effect. 

\begin{figure}[t]
    \begin{minipage}[t]{0.57\columnwidth}
        \centering
        \vspace{0pt}
        \includegraphics[width=\linewidth]{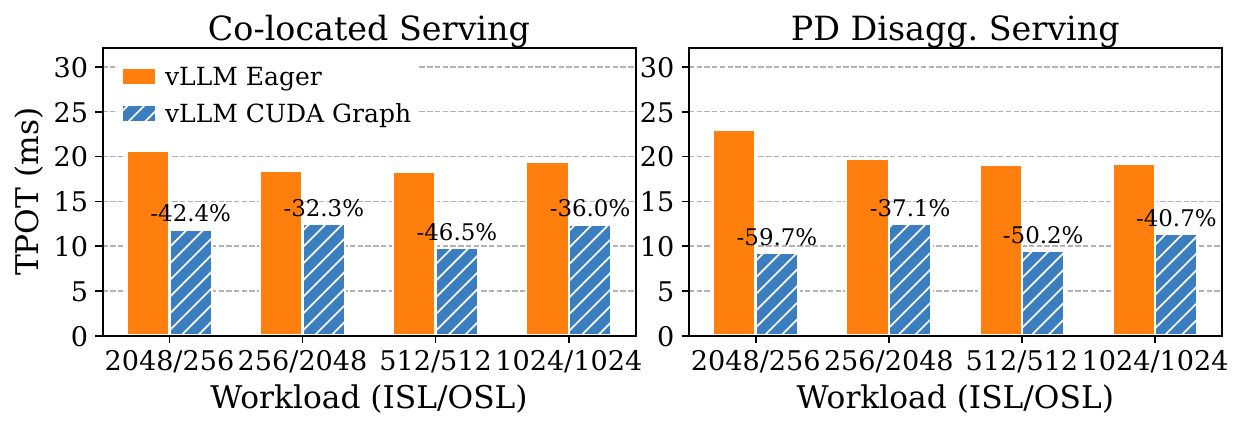}
        \vspace{-6mm}
        \caption{Measured vLLM TPOT with and without CUDA Graph under different workloads (64 requests per workload, mean ISL/OSL, tested on 8$\times$A800-SXM GPUs). Left: co-location. Right: PDD. Percentages show reduction.}
        \Description{A two-panel TPOT bar chart comparing vLLM CUDA Graph and vLLM eager across four workloads in co-location and PD disaggregation modes. Speedup percentages are annotated above the CUDA Graph bars.}
        \label{fig:motivation-c1-vllm-only}
    \end{minipage}%
    \hfill%
    \begin{minipage}[t]{0.40\columnwidth}
        \centering
        \vspace{0pt}
        \vspace{1.0mm}
        \small
        \resizebox{\linewidth}{!}{%
            \begin{tabular}{@{}llcc@{}}
            \toprule
            \textbf{Mode} & \textbf{ISL/OSL} & \textbf{Padding} & \textbf{Inflation} \\
            \midrule
            \multirow{4}{*}{\centering Co-location} & 2048/256  &   7K & 22.6\% \\
                                         & 256/2048  & 100K & 38.7\% \\
                                         & 512/512   &  29K & 45.8\% \\
                                         & 1024/1024 &  28K & 22.6\% \\
            \midrule
            \multirow{4}{*}{\centering PDD}  & 2048/256  &  14K & 42.6\% \\
                                         & 256/2048  & 111K & 42.5\% \\
                                         & 512/512   &  37K & 57.2\% \\
                                         & 1024/1024 &  52K & 40.0\% \\
            \bottomrule
            \end{tabular}}%
        \captionof{table}{CUDA Graph decode padding overhead.
        \emph{Padding}: wasted slots from batch alignment. 
        \emph{Inflation}: extra vs.\ useful tokens.}
        \label{tab:motivation-c1-padding}
    \end{minipage}
    \vspace{-4mm}
\end{figure}

\begin{figure}[t]
    \centering
    \includegraphics[width=\linewidth]{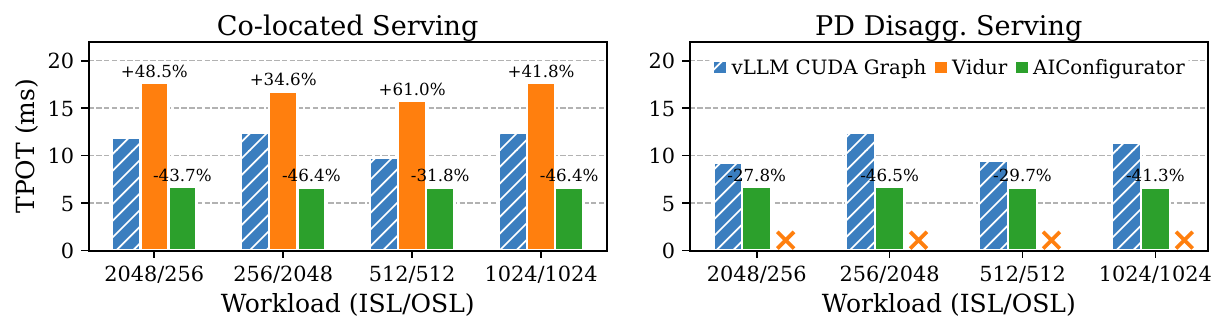}
    \vspace{-8mm}
    \caption{Existing simulators vs.\ vLLM CUDA Graph. Percentages represent error.}
    \Description{A two-panel TPOT bar chart comparing vLLM CUDA Graph and AIConfigurator across four workloads in co-location and PD disaggregation modes. Percentages above AIConfigurator bars show prediction error relative to vLLM CUDA Graph.}
    \label{fig:motivation-c1-baselines}
    \vspace{-4mm}
\end{figure}

\vspace{0.3em}
\noindent \textbf{Challenge 1:} \textit{How to model the complex runtime behaviors and control flows of advanced serving systems?}

Completeness first breaks at the \emph{serving-architecture level}.
% Modern LLM serving has moved beyond identical replicas running a monolithic engine: PDD separates compute-heavy prefill from memory-heavy decode and can serve up to $7.4\times$ more requests than co-located serving, while AFD further separates decode attention from FFN execution for large MoE models~\cite{stepfun2025step3largeaffordablemodelsystem,zhu2025megascale}.
Disaggregated serving architectures (PDD/AFD) can achieve up to a $7.4\times$ throughput improvement over traditional co-located serving~\cite{zhong2024distserve}. 
This leap is driven by a fundamental shift from monolithic execution to role-specific cluster decoupling~\cite{stepfun2025step3largeaffordablemodelsystem,zhu2025megascale}.
These architectures change the event graph itself: requests traverse role-specific clusters, KV-cache or activation transfers become causal edges, and MoE EP introduces routing-dependent synchronization.
Existing simulators, however, are typically built around a homogeneous-replica abstraction, where each replica owns the full model path, scheduler state, and local batch execution loop.
While sufficient for co-located serving, this abstraction cannot express PDD/AFD by simply adding a latency term; it requires role-specific workers, cross-cluster dependencies, and separate parallel domains as first-class execution objects.

\begin{figure}[t]
    \centering
    \includegraphics[width=\linewidth]{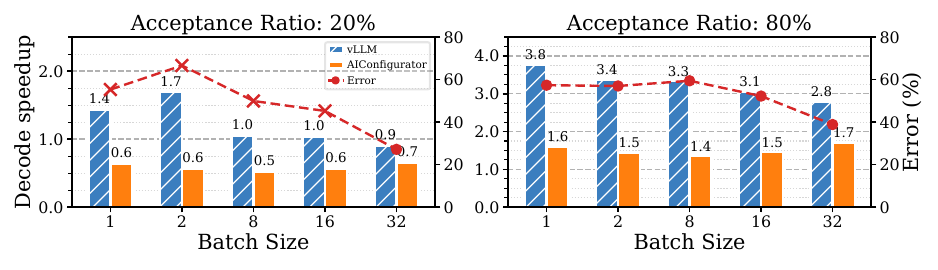}
    \vspace{-8mm}
    \caption{Qwen3-30B MoE MTP decode speedup ($\text{verify\_tokens}{=}4$). Bars (left axis): vLLM vs.\ AIConfigurator. Dashed line: prediction error. 
    Red $\times$: \emph{sign-mismatch} (measured speedup $>1$, predicted slowdown $<1$). 
    Ground-truth acceptance rates and fixed batch sizes are used to isolate other modeling errors.}
    \Description{A two-panel bar chart with an overlaid error line comparing vLLM and AIConfigurator decode speedup across batch sizes 1, 2, 8, 16, 32 at target acceptance ratios 0.2 and 0.8 with verify_tokens=4. Sign-mismatch points are highlighted with red x markers.}
    \label{fig:motivation-c1-mtp}
    \vspace{-0.5em}
\end{figure}

\begin{figure}[t]
    \centering
    \begin{minipage}[t]{0.46\linewidth}
        \centering
        \includegraphics[width=\linewidth]{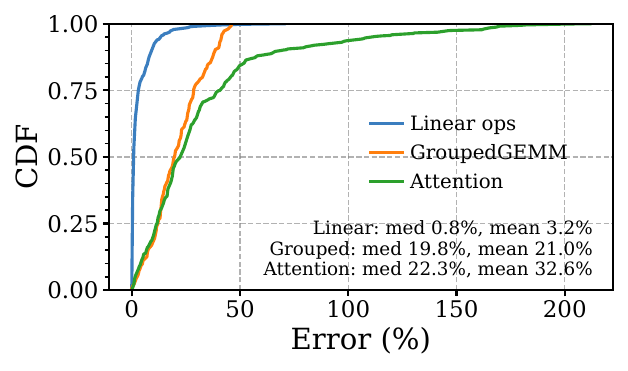}
    \end{minipage}
    \hfill
    \begin{minipage}[t]{0.5\linewidth}
        \centering
        % Align the compact bar plot with the CDF panel in the single-column layout.
        \raisebox{0.25cm}{\includegraphics[width=\linewidth]{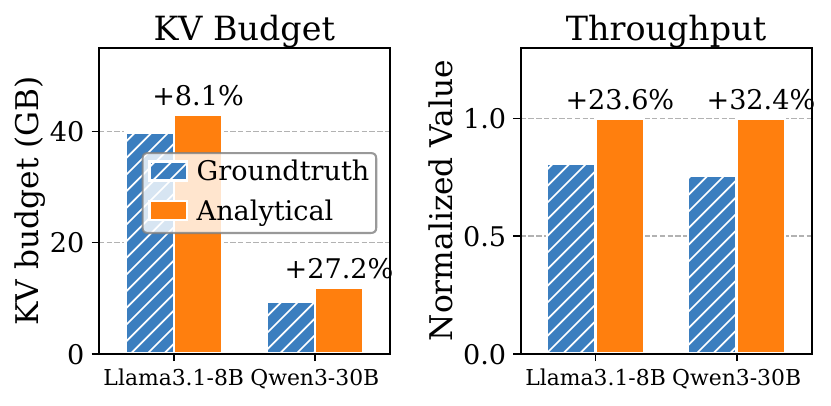}}
    \end{minipage}
    \vspace{-5mm}
    \caption{Fidelity gaps caused by simplified modeling.
             Left: Relying on coarse proxies (total token count) fails to capture batch heterogeneity, yielding coarse-grained performance estimates.
             Right: Analytical KV-cache modeling overestimates effective memory budget, leading to a cascade of errors from admission control to over-optimistic throughput projection.}
    \Description{A side-by-side figure. The left panel is a CDF showing prediction error for linear operators, GroupedGEMM, and attention. The right panel contains two compact bar charts comparing groundtruth and analytical memory budgets and normalized decode throughput for Llama-3.1-8B and Qwen3-30B MoE. Analytical budgeting overestimates memory budget and decode throughput.}
    \label{fig:motivation-c2-budget-throughput}
    \vspace{-0.2em}
\end{figure}

Completeness also breaks at the \emph{runtime-behavior level}.
Runtime optimizations are often treated as optional speedup knobs, but they change scheduler-visible state and batch execution semantics.
For example, vLLM captures decode CUDA Graphs at fixed batch-size bins $\{1,2,4,8,16,32,64\}$; a pure-decode step between bins is replayed at the next larger bin, so 33 active requests execute as a 64-slot graph.
As Figure~\ref{fig:motivation-c1-vllm-only} shows, CUDA Graph reduces decode TPOT by $32.3\%\mathrm{-}46.5\%$ under co-location and $37.1\%\mathrm{-}59.7\%$ under PDD, but its padding increases decode-token work by $22.6\%\mathrm{-}45.8\%$ and $40.0\%\mathrm{-}57.2\%$, respectively (Table~\ref{tab:motivation-c1-padding}).
This padding also changes runtime memory usage and KV-cache state for admission, chunking, and preemption.
Ignoring CUDA Graph therefore yields an eager-style execution path and the workload-insensitive TPOT surface in Figure~\ref{fig:motivation-c1-baselines}.
Analytical shortcuts fail similarly.
For speculative decoding, even with vLLM's per-step Qwen3-30B MoE multi-token prediction (MTP)~\cite{liu2024deepseek} acceptance rates\footnote{
Since AIConfigurator lacks native support for Qwen's MTP, we extend it by adapting its default MTP module to the Qwen MoE.}, an analytical MTP model deviates by $27.2\%\mathrm{-}66.6\%$ and predicts the opposite trend on $4$ of $5$ low-acceptance batches (Figure~\ref{fig:motivation-c1-mtp}).
Its scalar expectation formula flattens the non-linear speedup from event-driven variable commits, causing large cost overestimation under low acceptance rates.

\begin{figure}[t]
    \centering
    \includegraphics[width=0.85\linewidth]{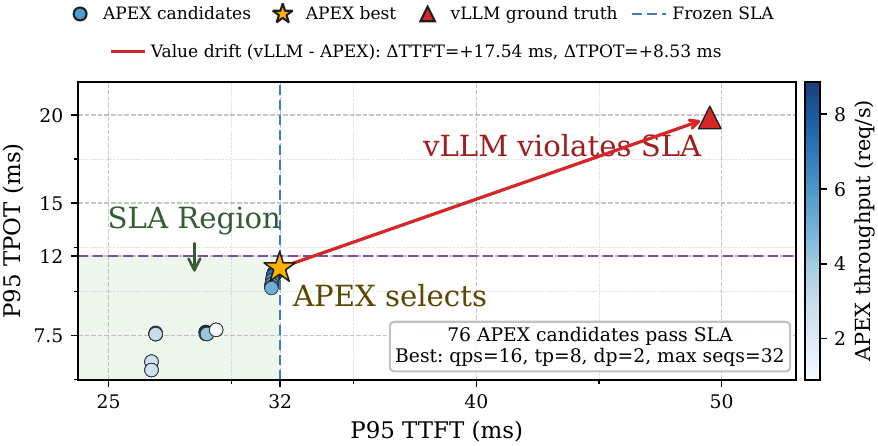}
    \vspace{-3mm}
    \caption{Decision drift from fidelity gaps on Llama-3.1-8B over 16 H800 GPUs (co-location).
    The simulator-selected best configuration lies inside the frozen SLA region,
    but the corresponding vLLM ground-truth point moves outside the SLA because overly optimistic comp-op and KV-cache budget predictions underestimate request latency.
    Note that we maintain feature parity in this evaluation by disabling vLLM optimizations unsupported by APEX.}
    \Description{A scatter plot of P95 TTFT and P95 TPOT. APEX candidates inside the frozen SLA region are shown as blue circles, the simulator-selected best point is shown as a star, and the vLLM ground-truth point is shown as a red triangle outside the SLA. A red arrow shows the drift from the simulator-selected point to the vLLM ground-truth point.}
    \label{fig:motivation-c2-frontier}
    % \vspace{-1em}
\end{figure}

\vspace{0.3em}
\noindent \textbf{Challenge 2:} \textit{How to accurately model the computation and memory dynamics within a closed-loop serving system?}

LLM inference serving is a closed-loop system: execution cost shapes memory state, memory constrains admission, and admission governs queueing and future batch composition.
Within this loop, modest per-event errors can accumulate into different operating regimes.
We find that fidelity breaks at two levels---per-operator runtime prediction and engine-level memory accounting---both of which are necessary for decision-grade simulation.

At the operator level, the key challenge is generalization: batch composition is determined online by the scheduler, making the execution space infeasible to profile exhaustively.
Existing simulators therefore rely on predictors fitted from primary operator profiles~\cite{agrawal2024vidur,xu2026aiconfiguratorlightningfastconfigurationoptimization,cho2026llmservingsim,lin2024apex}, using coarse proxy features such as total token count while ignoring per-request variance within a batch.
As shown in Figure~\ref{fig:motivation-c2-budget-throughput} (left), this leads to substantial errors under dynamic workloads; for example, Vidur's token-count predictor has a mean error of $32.6\%$ for FlashAttention~\cite{dao2024flashattention} ($0.151$\,ms predicted vs.\ $0.224$\,ms measured), with similarly large errors for MoE GroupedGEMM due to data-dependent router load distributions.
Such errors compound across layers and decode steps: even $0.1$\,ms per operator can add tens of milliseconds per request, exceeding the margins between competing serving configurations.

At the engine level, memory fidelity determines how the scheduler admits, chunks, and preempts requests.
The KV-cache budget is the residual GPU memory after allocating weights, activation buffers, CUDA Graph capture regions, and framework-internal overheads; approximating it as ``total minus weights'' shifts the scheduler's operating point.
In our scheduler-isolation replay (Figure~\ref{fig:motivation-c2-budget-throughput}, right), the analytical budgeting used by most existing simulators overestimates the available KV-cache budget by $8.1\%$ for Llama-3.1-8B and $27.2\%$ for Qwen3-30B MoE.
Since the scheduler uses this budget to bound concurrent decode slots, this initialization error inflates simulated decode throughput by $23.6\%$ and $32.4\%$, respectively---enough to alter capacity planning decisions.

Overall, per-component errors propagate through the closed-loop dynamics into decision-grade failures.
As Figure~\ref{fig:motivation-c2-frontier} shows, optimistic operator-runtime and KV-cache budget predictions can lead the simulator to select a ``best'' configuration that violates the SLA on the real system---precisely the failure mode simulation is meant to prevent.

%!TEX root = newmain.tex
%

\section{\sys Design}
\label{sec:design}

\noindent
\sys is a discrete-event simulator designed for LLM serving. It features a disaggregated abstraction that separates hardware-aware prediction from the control and execution planes.
This decomposition is grounded in a key observation: production serving
engines (e.g., vLLM~\cite{kwon2023efficient},
SGLang~\cite{zheng2024sglang}) share a common control-flow
skeleton---admission, batching, execution, completion---yet differ mainly
in the scheduling policies they apply and the operator backends they invoke.
\sys exploits this shared structure: it can integrate framework-specific
scheduling policies and operator backends without modifying the simulation core.

% \noindent\textbf{Key design choices.}
% \begin{itemize}[leftmargin=*]
%     \item \sys models \textbf{serving architecture as composable cluster
%     wiring} rather than hard-coded execution paths.  Co-location, PDD,
%     and AFD are constructed by composing the same cluster and replica
%     worker primitives with different transfer semantics; a new
%     architecture requires new wiring, not a new simulator
%     core~(\S\ref{subsec:topology_runtime}).

%     \item \sys treats \textbf{runtime optimizations as scheduling-state
%     transformations}.  CUDA Graph, speculative decoding, prefix caching,
%     and chunked prefill are modeled as adapters that intercept and
%     reshape events, batch composition, and memory state---changing which
%     events the engine produces next, not merely the current iteration's
%     latency~(\S\ref{subsec:topology_runtime}).

%     \item \sys provides a \textbf{Fidelity Plane} that
%     predicts per-operator runtime, communication latency, and memory
%     capacity from hardware profiles.  These predictions feed back into
%     scheduling as control inputs: compute cost determines event
%     completion and queueing; memory capacity governs admission and
%     preemption~(\S\ref{subsec:fidelity_stack}).
% \end{itemize}

\subsection{Simulator Overview}
\label{subsec:simulator_overview}

\begin{figure}[t]
    \centering
    \includegraphics[width=0.9\linewidth]{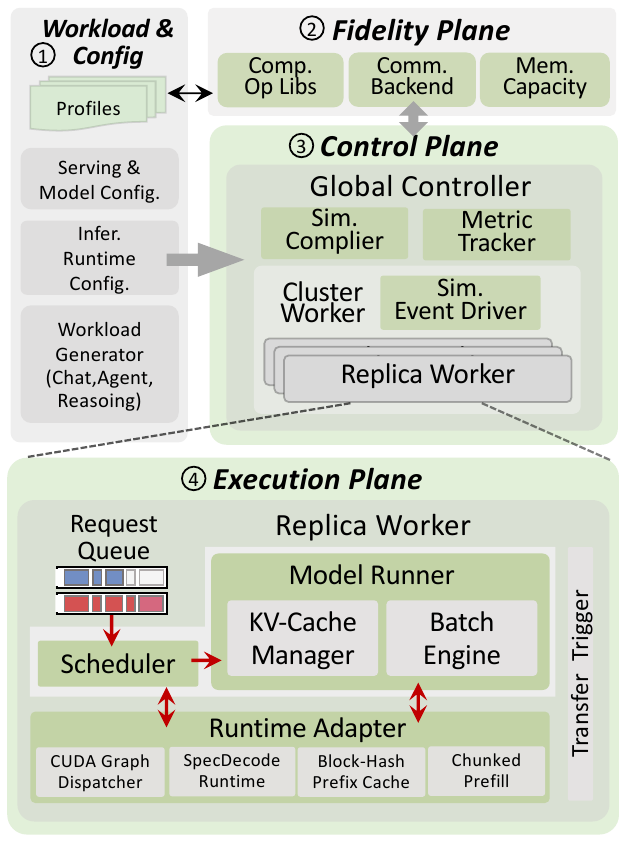}
    \vspace{-2.5mm}
    \caption{\sys system architecture.}
    \vspace{-4mm}
    \label{fig:archi}
\end{figure}

\noindent\textbf{System architecture.}
Figure~\ref{fig:archi} shows the system architecture.  \sys is organized into
four modules.
\ding{192}~\emph{Workload and Config.} It defines the serving scenario---model,
serving architecture, parallelism, runtime options, and request workloads.
\ding{193}~\emph{Fidelity Plan.} It provides calibrated per-operator,
communication, and memory-capacity (KV-cache budget) predictions that the \textit{execution plane}
queries on demand.
\ding{194}~\emph{Control Plane.} It provides disaggregation primitives and dependency-based control flow. It translates these specifications into a functional simulation and coordinates its execution.
\ding{195}~\emph{Execution Plane.} It enables feature-specific execution flows, advancing requests through scheduling, batching, runtime adaptation, and cross-cluster transfers.

\vspace{0.3em}
\noindent\textbf{System workflow.}
The workflow involves four steps.

\textit{Configuration and compilation.}
The user specifies the serving scenario---model, hardware, serving
architecture, parallelism, runtime features, and request
workloads---through CLI parameters, alongside calibrated profiles
consumed by the Fidelity Plane.  The Simulation Compiler in the
Control Plane reads these inputs and instantiates the simulation
topology.  Two abstractions map the simulation to a real serving
deployment: a \emph{Cluster Worker} represents a logical device group
serving a specific role (e.g., a prefill cluster, a decode cluster, or
a co-located cluster), and each contains multiple \emph{Replica
Workers}, each of which models a serving replica spanning GPUs
under the configured parallelism.

\textit{Event-driven simulation loop.}
Each Cluster Worker is driven by a per-cluster discrete-event engine
that maintains a priority event queue and advances events in timestamp
order.  For disaggregated serving architectures, the per-cluster
drivers run in parallel (one thread per cluster) and coordinate
through inter-cluster event queues.  Once initialization completes,
the workload generator injects request arrival events into the cluster queues, triggering the simulation loop.

\textit{Per-iteration request processing.}
When a request arrival fires, the Replica Worker's Scheduler decides
admission and batch composition; Runtime Adapters then reshape the
batch before execution~(\S\ref{subsec:execution_plane}).  The Batch
Engine queries the Fidelity Plane for per-operator runtime predictions
(both computation and communication), producing the iteration's
completion time~(\S\ref{subsec:fidelity_plane}).  For disaggregated
systems, batch completion triggers a cross-cluster transfer event; the
downstream cluster resumes processing only after the transfer
completes.

\textit{Metrics and output.}
A Metric Tracker records each request's history across its entire
lifecycle---spanning admission, batching, iteration execution, and
cross-cluster transfers.  When the simulation finishes, the simulator
outputs per-request latencies, batch traces, TTFT/TPOT breakdowns,
throughput, E2E makespan, and memory utilization reports.

\vspace{-1em}
\subsection{Control Plane}
\label{subsec:control_plane}
The Control Plane compiles a user-level serving specification into a runnable simulation by instantiating workers, binding communication domains, wiring cross-stage event dependencies, and setting Execution Plane capacity constraints.

\noindent\textbf{Input compilation.}
The Simulation Compiler resolves the model, hardware, serving architecture,
runtime features, workload generator, and backend choices into typed
simulation objects.  Co-location yields a monolithic cluster; PDD
yields prefill and decode clusters; AFD splits decode into attention
and FFN (MoE) clusters.
Table~\ref{tab:simulator-config} lists two
representative configurations.

\noindent\textbf{Parallel primitives.}
Production engines expose parallelism through different and often
implicit vocabularies; a simulator that compiles all of them onto a
single backend needs a parallelism expression that is both
\emph{explicit}---no coupling hidden inside the runtime---and
\emph{expressive} enough to subsume every framework's sharding
semantics without ambiguity. \sys introduces a two-domain
decomposition for this purpose: a replica is parameterized by the
pipeline size $\mathrm{pp}$ together with an attention sharding
tuple
$(\mathrm{tp}_{\mathrm{attn}}, \mathrm{dp}_{\mathrm{attn}})$ and an
FFN sharding tuple
$(\mathrm{tp}_{\mathrm{ffn}}, \mathrm{ep}_{\mathrm{ffn}})$, where
$\mathrm{ep}_{\mathrm{ffn}}$ degenerates to $\mathrm{dp}_{\mathrm{ffn}}$
on dense models.  This decomposition is the unit at which the Fidelity Plane
and the communication backend are queried.
Each serving architecture maps these two domains onto a distinct set
of \textit{cluster roles}, denoted in sans-serif throughout:
co-location uses a single cluster $\mathsf{C}$ hosting both domains;
PDD uses a prefill cluster $\mathsf{P}$ and a decode cluster
$\mathsf{D}$, each independently hosting both domains; AFD uses a prefill cluster $\mathsf{P}$, an attention cluster $\mathsf{A}$, and an FFN cluster $\mathsf{F}$, each
hosting a single domain.  Whenever a role $c$ hosts both domains,
the two sharding products must span the same device set:
\vspace{-0.3em}
\begin{equation}
\mathrm{tp}_{\mathrm{attn}} \cdot \mathrm{dp}_{\mathrm{attn}}
\;=\;
\mathrm{tp}_{\mathrm{ffn}} \cdot \mathrm{ep}_{\mathrm{ffn}},
\quad
c \in \{\mathsf{C},\, \mathsf{P},\, \mathsf{D}\}.
\label{eq:shared_domain}
\end{equation}
The per-replica world size of each cluster role is then
\begin{equation}
\vspace{-0.5em}
W_{R}^{\,c}
\;=\;
\begin{cases}
\mathrm{pp} \cdot \mathrm{tp}_{\mathrm{attn}} \cdot \mathrm{dp}_{\mathrm{attn}},
  & c \in \{\mathsf{C},\, \mathsf{P},\, \mathsf{D},\, \mathsf{A}\}, \\[2pt]
\mathrm{pp} \cdot \mathrm{tp}_{\mathrm{ffn}} \cdot \mathrm{ep}_{\mathrm{ffn}},
  & c = \mathsf{F},
\end{cases}
\vspace{-0.5em}
\label{eq:wr_per_role}
\end{equation}
where the two branches agree on $\mathsf{C}/\mathsf{P}/\mathsf{D}$ by
Eq.~\ref{eq:shared_domain}.  Each role $c$ instantiates $N_{R}^{\,c}$
replicas (the deployment pod count), so the total simulated system
size is $\sum_{c} N_{R}^{\,c} \cdot W_{R}^{\,c}$.  At initialization,
every active parallel axis is registered as a domain group with the
Comm.\ Backend, so that a later collective query (attention
all-reduce, FFN all-to-all, etc.) resolves to a backend-native group
of the correct shape.

% \noindent\textit{Example: expressing vLLM's parallelism.}  A vLLM
% co-location runtime configured with
% $(\mathrm{pp}, \mathrm{tp}, \mathrm{dp}, \mathrm{ep})$ maps onto the
% \sys primitives as
% $\mathrm{tp}_{\mathrm{attn}}{=}\mathrm{tp}$,
% $\mathrm{dp}_{\mathrm{attn}}{=}\mathrm{dp}$,
% $\mathrm{tp}_{\mathrm{ffn}}{=}1$,
% $\mathrm{ep}_{\mathrm{ffn}}{=}\mathrm{ep}$, and
% $\mathrm{pp}{=}\mathrm{pp}$.  Substituting into
% Eq.~\ref{eq:shared_domain} yields
% $\mathrm{tp}\cdot\mathrm{dp} = \mathrm{ep}$, which is exactly the
% implicit EP-sizing constraint that vLLM never states but always
% enforces internally---surfaced here as an explicit and checkable
% invariant of the \sys parallelism expression.  Other engines
% (SGLang, Megatron-LM, etc.) map by filling in the same five
% primitives.

\begin{table*}[!t]
\centering
\normalsize
\setlength{\tabcolsep}{2.6pt}
\renewcommand{\arraystretch}{1.18}
\resizebox{\textwidth}{!}{%
\begin{tabular}{|c|c|c|c|c|c|c|c|c|c|c|c|c|c|c|c|c|c|c|c|c|c|c|}
\hline
\multicolumn{2}{|c|}{\textbf{Model}} &
\multicolumn{2}{c|}{\textbf{Hardware}} &
\multicolumn{3}{c|}{\textbf{Engine Backend}} &
\multicolumn{5}{c|}{\textbf{Serving Config}} &
\multicolumn{1}{c|}{\makecell{\textbf{Chunked}\\\textbf{Prefill}}} &
\multicolumn{2}{c|}{\makecell{\textbf{Speculative}\\\textbf{Decoding}}} &
\multicolumn{7}{c|}{\textbf{Other Runtime Features}} &
\textbf{...} \\
\hline
\textbf{Name}
& \makecell{\textbf{Param.}\\\textbf{Size}}
& \textbf{GPU}
& \textbf{Topo.}
& \textbf{Runtime}
& \makecell{\textbf{Attn.}\\\textbf{Backend}}
& \makecell{\textbf{Comm.}\\\textbf{Backend}}
& \makecell{\textbf{Sys.}\\\textbf{Arch.}}
& \makecell{\textbf{Parallel}\\\textbf{Param.}}
& \makecell{\textbf{Batch Size}\\\textbf{Cap}}
& \makecell{\textbf{Max Batch}\\\textbf{Tokens}}
& \makecell{\textbf{GPU Mem.}\\\textbf{Util.}}
& \makecell{\textbf{Token}\\\textbf{Thresh.}}
& \makecell{\textbf{Accept}\\\textbf{Ratio}}
& \makecell{\textbf{Verify}\\\textbf{Tokens}}
& \makecell{\textbf{MoE}\\\textbf{Routing}}
& \makecell{\textbf{CUDA}\\\textbf{Graph}}
& \makecell{\textbf{Prefix}\\\textbf{Cache}}
& \makecell{\textbf{Thinking}\\\textbf{Mode}}
& \makecell{\textbf{Tool}\\\textbf{Call}}
& \textbf{Quant.}
& \makecell{\textbf{Hier.}\\\textbf{Cache}}
& \textbf{...} \\
\hline
Llama 3.1 & 405B
   & \makecell{512$\times$H100\\512$\times$H20} & Fat-tree
   & vLLM & FlashInfer & HTSim
   & PDD
   & \makecell{P: pp16/tp8/dp2\\D: pp16/tp8/dp6}
   & 256 & 16K & 0.90
   & 2K
   & 30\% & 16
   & Balanced
   & \ding{51} & \ding{51} & -- & -- & FP8 & \makecell{CPU\\offload} & ... \\
\hline
Qwen3 & 235B
   & \makecell{1024$\times$H800\\1024$\times$H20} & Fat-tree
   & SGLang & Triton & ASTRA-Sim
   & AFD
   & \makecell{P: pp8/tp8/dp8/ep64\\A: pp8/tp8/dp4\\F: pp8/tp8/dp32/ep256}
   & 512 & 8K & 0.85
   & 1K
   & 25\% & 4
   & Skew
   & \ding{51} & \ding{51} & depth 4 & \makecell{p95\\200ms} & FP16 & \makecell{CPU\\offload} & ... \\
\hline
\end{tabular}%
}
\caption{Representative simulator configurations. Column names follow the corresponding \sys configuration fields.}
\label{tab:simulator-config}
\vspace{-4mm}
\end{table*}

\vspace{0.2em}
\noindent\textbf{Dependency definition.}
The Control Plane materializes five classes of cross-stage dependency
as explicit events, so that queueing and
synchronization follow directly from the event graph.
\textit{(i)~Pipeline stages.}
Within a replica, the $pp$ stages form a strict chain: a stage-end
event on stage $s$ schedules stage $s{+}1$, with one microbatch of lag
between adjacent stages.
\textit{(ii)~PDD KV-cache transfer.}
Prefill completion of a request emits a \texttt{KVCacheTransferStart}
event to the decode cluster; the matching \texttt{End} event admits the
request into decode scheduling.  Transfer latency is queried from the
Fidelity Plane using KV-cache size, link topology, and current transfer
concurrency.
\textit{(iii)~AFD activation transfer.}
Every decode iteration emits paired \texttt{M2NTransferStart}/
\texttt{End} events between the attention and FFN clusters.
\textit{(iv)~MoE EP synchronization.}
For every MoE layer, \sys emits per-rank
\texttt{AllToAllCombineReady} events and only fires the combine
collective once the slowest rank has reported ready; straggler waits
therefore arise from per-rank prediction rather than from a separate
imbalance model.
In the Appendix~(\cref{app:pdd-afd-workflow}), we detail the workflows and dependencies
of disaggregated architectures (PDD and AFD).
% This contrasts with AIConfigurator~\cite{} and Apex~\cite{},
% both of which support EP topology but assume balanced expert load and
% so cannot reproduce routing-induced stragglers.

\noindent\textbf{Agentic Reasoning.}
\sys introduces a \emph{stateful request abstraction} for
new workloads such as reasoning, tool-using agents, and RL rollouts.
The Control Plane treats multi-round thinking/reasoning as a per-request event chain.
Under reasoning, a request carries (a) a number of
\emph{reasoning rounds} to execute, (b) a \emph{tool-execution
latency} attached to each inter-round tool call, and (c) a per-round
plan of prefill/decode token counts.  Each intermediate round runs
as an ordinary prefill$\to$decode cycle; on completion it emits a
\texttt{ThinkingRequeue} event that re-admits the request after
the tool-execution delay, with session affinity to the same
replica so that the KV cache populated in the previous round can be
served as a prefix-cache hit. The final round produces the
user-facing response.

\noindent\textbf{Runtime invariants.}
Before any batch is formed, the Control Plane fixes the capacity
envelope used during execution (details in~\cref{subsec:fidelity_plane}).
% : per-replica KV-block budget,
% request-slot limits, and FFN activation capacity, computed after
% subtracting weight memory, non-KV overhead, memory margin, and
% runtime reservations.  
It also records runtime contracts later used by admission, chunking, and preemption, including CUDA Graph capture sizes, prefix-cache eligibility, and speculative-decoding token allowances.
Memory safety is enforced both at initialization and during execution: the KV-cache Manager aborts launch if weights do not fit or the resolved KV-cache block count is zero, while the Scheduler checks block availability against a watermark before each admission and triggers preemption instead of overcommitting capacity.

\subsection{Execution Plane}
\label{subsec:execution_plane}
The Execution Plane advances the request lifecycle within each
replica: admission, batch composition, and per-iteration cost
resolution.
It schedules workloads under control-plane constraints and supports runtime optimization via an adapter.
% It is where the event graph wired by the Control
% Plane is stepped through in simulated time.

\noindent\textbf{Scheduler.}
\sys implements the Scheduler as a standalone module, following production engines such as vLLM~\cite{kwon2023efficient} and SGLang~\cite{zheng2024sglang}, while keeping their scheduling logic unchanged.
Only the I/O layer is rewired---
the request queue is fed by simulated arrivals, the batches are
consumed by the DES event loop, and memory
pressure is read from a simulated KV-cache block counter.  Preserving
the mechanism is what lets \sys reproduce engine-specific behavior
under pressure (e.g., vLLM watermark preemption~\cite{watermark}, SGLang's
admission rules~\cite{admission}).
We provide additional simulation results on large-scale clusters
comparing SGLang- and vLLM-v1-style schedulers in the Appendix~(\cref{app:sglang-vllm-scheduler}).

\noindent\textbf{Runtime adapters.}
\sys carves out Runtime Adapters as an abstraction
through which advanced engine features are supported.
An adapter is a feature-specific rule attached to the Scheduler--Batch
Engine loop.
Each adapter declares how one optimization changes a
well-defined slice of the loop: (i)~the \emph{scheduler-visible
state} (e.g., which prompt blocks count as already computed),
(ii)~the \emph{batch shape} the Batch Engine queries the Fidelity Plane
with (e.g., the padded size under CUDA Graph), or (iii)~\emph{per-request progress} (e.g., how many decode tokens commit per step).
We describe two in detail; the remaining ones follow the same pattern.

\textit{\textbf{CUDA Graph.}}
Captured graphs remove per-kernel launch overhead only for captured batch shapes.
The adapter pads each decode batch to the next larger captured size, then queries the Fidelity Plane with the kernel-only measurement family provided by the computation operator lib (\cref{subsec:fidelity_plane}).
Non-graph executions use the launch-inclusive family instead.
This models CUDA Graph's coupled effects: padding increases compute work, while graph replay removes launch overhead.
A scalar speedup factor cannot capture both.

\textit{\textbf{Speculative decoding (MTP).}}
The adapter maintains \emph{per-request} state (planned, verified,
accepted, committed token counts).
For MTP, each decode step becomes a draft$\to$verify
$\to$commit cycle in which the verify phase is a prefill-like
forward pass that can share a batch with ordinary decode work.
Per-request accounting lets different requests within the same
batch carry different speculative depths and different acceptance
outcomes, which is a prerequisite for capturing the batch-shape
variance that shapes tail latency under speculative workloads.

% \noindent\textit{Prefix cache and chunked prefill.}
% Prefix caching is modeled by a block-hash index that marks
% matched prefix blocks as already computed without storing any KV
% tensor data.  Chunked prefill partitions the prefill token budget
% across waiting requests and carries un-chunked residuals to the
% next iteration.  Both are single-rule adapters that reuse the
% same "reshape batch, then query cost" pattern.
Other runtime features use the same interface.  Prefix caching marks
matched block hashes as already computed before admission and updates
the cache when full blocks complete; chunked prefill caps long-prompt
progress by the iteration token budget.

\subsection{Fidelity Plane}
\label{subsec:fidelity_plane}
The Fidelity Plane is a runtime-estimation layer queried by the Execution Plane for every batch iteration and cross-cluster transfer.
It has three components: a Compute Operator Library that predicts per-op runtime, a Memory-Capacity Model that enforces batch-admission constraints, and a Comm. Backend that estimates collective-communication costs.

\noindent\textbf{Compute operator library.}
\sys classifies every modeled operator by what determines its runtime and selects a tailored prediction strategy for each class:

\begin{itemize}[leftmargin=*, labelsep=1.5mm]
\vspace{-0.3em}
    \item[(i)] \textbf{\emph{Token-count operators}} (linear op family, e.g., GEMM, element-wise, norm). 
    Their runtime is primarily determined by the number of tokens processed within the operator's effective TP slice. 
    The feature set reduces to \texttt{num\_tokens}, with the TP slice resolved based on whether the operator is replicated or sharded. For this class, \sys fits either a linear-regression or a small random-forest predictor, which is configurable per operator.

    \item[(ii)] \textbf{\emph{Sequence-dependent operators}} (attention op family). 
    Their runtime varies non-linearly with the per-request sequence-length distribution inside a batch, rather than just aggregate totals. 
    \sys employs random-forest regressors over richer feature vectors. The features include aggregate counts (batch size, total tokens) and distributional statistics of per-request lengths (min, max, and percentiles of prefill and decode lengths). This design captures how kernel partitioning and tile scheduling react non-linearly to length heterogeneity, addressing why single length regressions (e.g., Vidur) diverge from measured latency under chunked prefill.

    \item[(iii)] \textbf{\emph{Routing-dependent operators}} (MoE op family). 
    Their runtime heavily depends on how tokens are dynamically assigned to experts. 
    Handled similarly by random-forest regressors, the feature vectors for MoE ops combine load-balance statistics (variance and max of token-to-expert counts), expert-selection ratio, expert count, and model dimensions. This ensures that workload skew at the expert level is faithfully preserved rather than averaged away.
\vspace{-0.3em}
\end{itemize}

We use separate profiling and training subsystems for these predictors.
Profiling runs operators on real GPUs in a \emph{single-GPU sharded} mode.
It materializes each per-rank slice locally and stubs out collectives, making collection independent of simulated cluster scale.
Training then fits the predictors.
Each kernel is measured in two modes: \emph{kernel-only} for GPU-side duration and \emph{launch-inclusive} for host-side launch overhead.

\vspace{0.2em}
\noindent\textbf{Memory capacity.}
\sys estimates the per-replica KV-cache block budget that admission
control relies on.  It does so through a vLLM-style dummy profile
run: the per-rank sharded model is instantiated with dummy weights
and executed for one sized forward pass, and the PyTorch allocator
snapshot yields three quantities---weight memory, torch peak
increase, and non-torch residency (e.g., NCCL workspace
buffers~\cite{NCCLbuffer})---all measured on a single representative rank.  Because
the sharded forward runs in a single-GPU process, simulating a
thousand-GPU deployment requires only one GPU to obtain the per-
rank footprint; under pipeline parallelism, \sys takes the
worst-case slice across PP stages as the representative rank.
The KV-cache Manager then derives the KV-cache block budget by subtracting
parameter memory (computed analytically from the weight shard) and
the profiled non-KV-cache overhead (torch peak increase plus non-torch
residency, beyond weights) from the requested GPU budget.  The
Scheduler (\S\ref{subsec:execution_plane}) consumes this budget for
admission, chunked-prefill partitioning, and watermark-triggered
preemption.

\vspace{0.2em}
\noindent\textbf{Comm. backend.}
\sys integrates existing network simulation tools (ASTRA-Sim~\cite{won2023astra} and HTSim~\cite{htsim}) as pluggable backends. 
It dynamically selects the backend based on the domain scale to balance simulation accuracy and cost.
% Collectives are routed through a pluggable backend interface that
% exposes the six standard primitives
% (\texttt{allreduce}, \texttt{allgather}, \texttt{broadcast},
% \texttt{send\_recv}, \texttt{reduce\_scatter}, \texttt{all\_to\_all})
% over the parallel-domain groups bound by the control plane.  A
% cluster may select Vidur, an analytical model, AstraSim,
% Collective-sim, or AIConfigurator as its backend; the choice is
% independent per cluster and does not touch the simulator core.

The Fidelity Plane composes flexibly with Runtime Adapters in the Execution Plane to support a variety of runtime optimizations.
For CUDA Graph, the adapter switches the Fidelity Plane's measurement mode (kernel-only vs.\ launch-inclusive) based on whether the current batch shape hits a capture bin, accurately reflecting the captured graph's speedup.
For chunked prefill, the Fidelity Plane's profiling subsystem supplies per-operator performance data across input shapes parameterized by chunk length, ensuring its performance impact is modeled correctly.
% \vspace{-1em}

% \input{benchmark_exploration}
% \input{comm_pred}
% \input{overlap_pred}
%!TEX root = newmain.tex

\section{Implementation}
\label{sec:implementation}

\sys is implemented in about 70K LoC of Python code, retaining Vidur's~\cite{agrawal2024vidur}
foundational modules and DES utilities while refactoring the
simulator out of its monolithic-replica execution path.  The core engineering
work is a cluster- and global-level execution stack: role-specific cluster
workers, global orchestration, cross-cluster event queues, and concrete paths
for KV-cache transfer, M2N transfer, comm. backend calls, and runtime adapters.
In addition, the \emph{Sim.\ Compiler} ingests model definitions in the
HuggingFace config format~\cite{huggingface}.
The \textit{compute operator library} exposes a uniform predictor interface
calibrated against the runtime APIs of mainstream serving stacks, covering 
PyTorch~\cite{paszke2019pytorch}, vLLM, and FlashInfer~\cite{ye2025flashinfer}; 
per-operator predictors are fitted once and reused across deployments. 
The \textit{comm. backend} is realized as a pluggable adapter behind a collective interface, with implementations over
ASTRA-Sim~\cite{won2023astra} and HTSim~\cite{htsim}, selectable per cluster without touching the simulator core.
\sys surfaces 808 typed CLI options organized into four orthogonal groups---workload, hardware, serving and runtime, and system metrics.

%!TEX root = newmain.tex

\section{Evaluation} 
\label{sec:evaluation}

\noindent\textbf{Setup.} Experiments are conducted on two servers, each hosting 8$\times$H800-SXM GPUs (400 GB/s NVLink, 400 Gb/s NDR IB per GPU). 
For fidelity evaluations, we use two MoE models---Qwen3-30B MoE~\cite{qwen3-a30b} and Step3-316B~\cite{stepfun2025step3largeaffordablemodelsystem} (MFA, 56 layers, 48+1 experts, top-3 routing)---and one dense model, Llama3.1-8B~\cite{llama3.1-8b}. 
We use vLLM v0.10.2 with the V1 engine as the ground truth.\footnote{Due to the absence of a publicly available AFD implementation, we use an in-house implementation for AFD calibration.}
We evaluate four online workload patterns by input sequence length (ISL) and output sequence length (OSL): (1) prefill-heavy (2048/256); (2) decode-heavy (256/2048); (3) balanced (1024/1024); and (4) a SharedGPT trace~\cite{SharedGPT}. 
System performance is assessed with request-level TTFT and TPOT (reported at P95 unless stated otherwise), throughput, and workload-level E2E makespan.
% When a table reports request-level end-to-end latency, we label it explicitly as E2E p95; E2E makespan denotes the completion window of the full replay and is not percentile-based.
We provide fidelity results for other GPU types (H20) in the Appendix~(\cref{app:h20-op-fidelity,app:h20-e2e-fidelity}).

\noindent\textbf{Baselines.} 
We compare \sys with four open-source simulators as mentioned in~\cref{sec:motivation}:
\begin{itemize}[leftmargin=3mm]
    \vspace{-0.2em}
    \item AIConfigurator~\cite{xu2026aiconfiguratorlightningfastconfigurationoptimization}: The SOTA simulator from NVIDIA, combining an op database with an analytical workflow.
    \item Vidur~\cite{agrawal2024vidur}: Microsoft's discrete-event simulator, lacking support for MoE and disaggregated architectures.
    \item LLMServingSim2.0~\cite{cho2026llmservingsim}: A simulator primarily focused on heterogeneous hardware systems.
    \item Apex~\cite{lin2024apex}: An emulator searching for optimal plans, lacking support for disagg. architectures and EP (limited).
\end{itemize}

\subsection{Operator and Mem. Fidelity}
\label{subsec:op_fidelity}

\begin{figure}[t]
    \centering
    \includegraphics[width=1\linewidth]{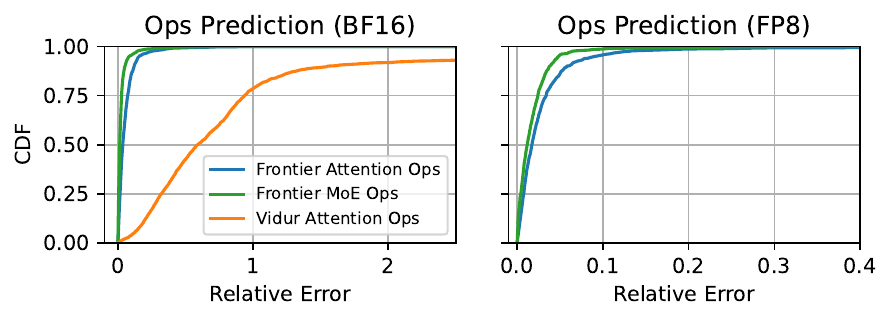}
    \vspace{-5mm}
    \caption{Per-operator relative-error CDF on H800 for attention, and MoE kernels under BF16 (left) and FP8 (right). Vidur is undefined for MoE ops, and its baseline cannot be evaluated under FP8.}
    \label{fig:op_cdf}
    \vspace{-2mm}
\end{figure}

\begin{figure}[t]
    \begin{minipage}[t]{0.45\columnwidth}
        \centering
        \vspace{0pt}
        \includegraphics[width=0.95\linewidth]{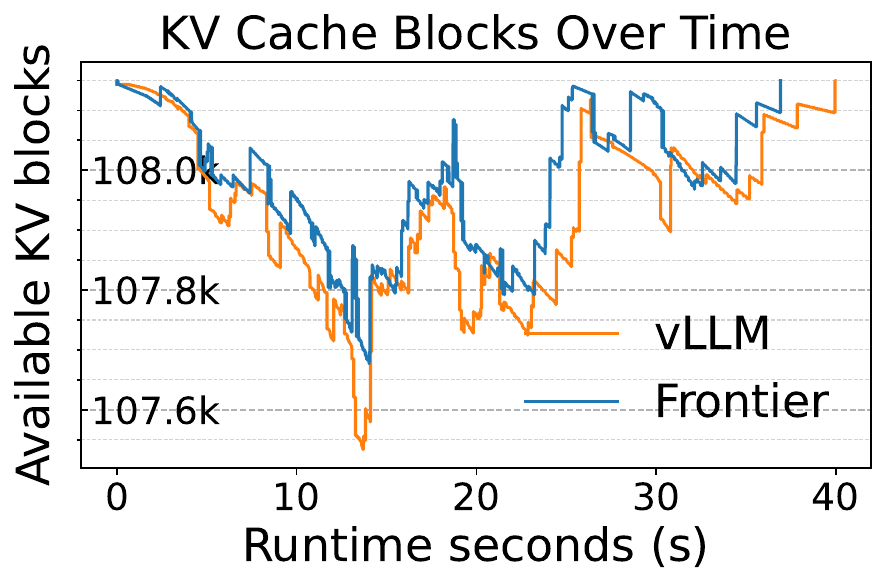}
        \vspace{-5mm}
        \caption{Available KV-cache blocks over time under co-location (Qwen3-30B MoE, SharedGPT trace), with a max gap of 294 blocks (115.6 MB; $\Delta\mathrm{MB}=0.393\Delta B$, where $\Delta B$ is the block gap).}
        \label{fig:moe_colocation_sharedgpt_kv_budget_timeline}
    \end{minipage}%
    \hfill%
    \begin{minipage}[t]{0.53\columnwidth}
        \centering
        \vspace{0pt}
        \vspace{1.5mm}
        \tiny
        \setlength{\tabcolsep}{1.2pt}
        \renewcommand{\arraystretch}{0.6}
        \resizebox{\linewidth}{!}{%
        \begin{tabular}{@{}ccccc@{}}
            \toprule
            Mode & Parallel & vLLM & $\Delta$\sys & $\Delta$Analytical \\
            \midrule
            \multirow{2}{*}{Co-loc.} & (1,8,1,8) & 31k & +7 (0.02\%) & +4.4k (14.10\%) \\
                    & (4,2,1,2) & 58.0k & +1.0k (1.76\%) & +12.5k (21.38\%) \\
            \midrule
            \multirow{2}{*}{Disagg.}   & (2,2,2,4) & 27.0k & +0.5k (1.89\%) & +7.6k (27.95\%) \\
                    & (1,4,1,4) & 12.0k & -1 (0.01\%) & +4.6k (39.73\%) \\
            \bottomrule
        \end{tabular}}%
        \captionof{table}{Initial KV-cache block budget accuracy for Qwen3-30B MoE. Parallel: (PP,TP,DP,EP). \sys uses profiled runtime non-KV-cache memory overhead in the KV-cache manager, while the analytical baseline omits the runtime profile. Percentages are relative errors to vLLM.}
        \label{tab:initial_kv_budget_profiled}
    \end{minipage}
    \vspace{-5mm}
\end{figure}

\noindent\textbf{Operator accuracy.}
Figure~\ref{fig:op_cdf} reports the relative-error CDF of per-operator latency prediction on H800 under BF16 and FP8, covering the attention, and MoE GroupedGEMM kernels that dominate the batch runtime.
Under BF16, \sys reaches p50/p95 errors of $3.5\%/14.2\%$ on attention, $3.3\%/6.4\%$ on linear, and $1.4\%/5.3\%$ on GroupedGEMM, whereas Vidur's token-only attention predictor, fit on the same population, reaches a p50/p95 of $55.4\%/376.1\%$ and is undefined for GroupedGEMM; FP8 shifts the \sys attention p95 further down to $8.8\%$, a regime Vidur does not support.
The residual is consistent with the classification in \S\ref{subsec:fidelity_plane}: attention latency is shaped by the per-request length distribution inside a batch and GroupedGEMM by expert-level routing load, so a token-aggregate regressor cannot separate compositions that diverge at the kernel's partitioning and tile-scheduling level.
% The tail of \sys is not closed: on the heterogeneity-stress stratum the attention p95 inflates to $31.6\%$, so severe length skew still exceeds the interpolation range of the current feature set and motivates broader coverage of these strata in the profiling subsystem.

\begin{figure}[t]
    \begin{minipage}[t]{0.45\columnwidth}
        \centering
        \vspace{0pt}
        \includegraphics[width=\linewidth]{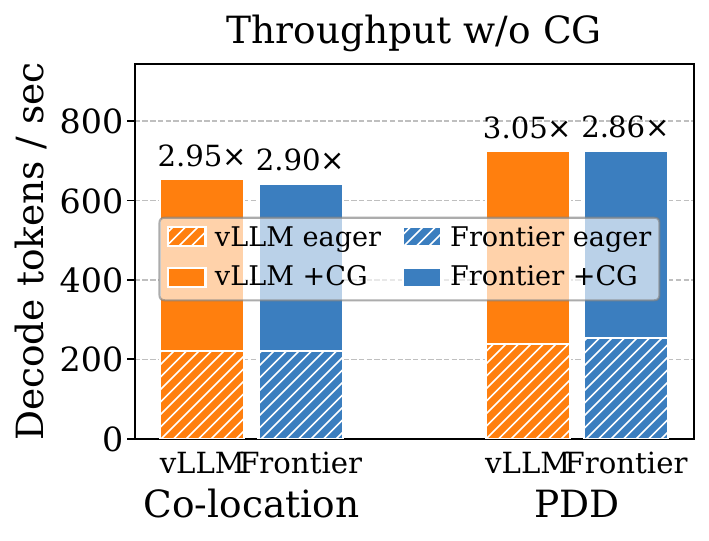}
        \vspace{-9mm}
        \caption{Decode throughput. +CG bars stack eager throughput with the CUDA Graph gain; top labels show total +CG/eager speedup.}
        \label{fig:cuda_graph_sharedgpt_speedup}
    \end{minipage}%
    \hfill%
    \begin{minipage}[t]{0.5\columnwidth}
        \centering
        \vspace{0pt}
        \vspace{7mm}
        \tiny
        \setlength{\tabcolsep}{1.2pt}
        \renewcommand{\arraystretch}{0.6}
        \resizebox{\linewidth}{!}{%
        \begin{tabular}{@{}ccccc@{}}
            \toprule
            Arch. & Workload & E-vLLM & CG-vLLM & $\Delta$\sys \\
            \midrule
            \multirow{3}{*}{Co-loc.} & Prefill-heavy & 294.8k & 307.3k & -6.9k (2.25\%) \\
                    & Decode-heavy & 294.8k & 420.5k & -10.7k (2.55\%) \\
                    & SharedGPT & 69.6k & 84.2k & +0.3k (0.40\%) \\
            \midrule
            \multirow{3}{*}{Disagg.} & Prefill-heavy & 147.5k & 150.1k & -1.2k (0.80\%) \\
                    & Decode-heavy & 147.5k & 167.9k & +0.6k (0.37\%) \\
                    & SharedGPT & 34.9k & 42.9k & -30 (0.07\%) \\
            \bottomrule
        \end{tabular}}%
        \captionof{table}{Compute-participating token accounting on Qwen3-30B MoE. E: eager; CG: CUDA Graph. Eager $\Delta$\sys is +0 (0\%) for all rows; CG $\Delta$\sys is the signed delta to vLLM.}
        \label{tab:cuda_graph_token_accounting}
    \end{minipage}
    \vspace{-5mm}
\end{figure}

\begin{figure}[t]
    \begin{minipage}[t]{0.54\columnwidth}
        \centering
        \vspace{0pt}
        \includegraphics[width=\linewidth]{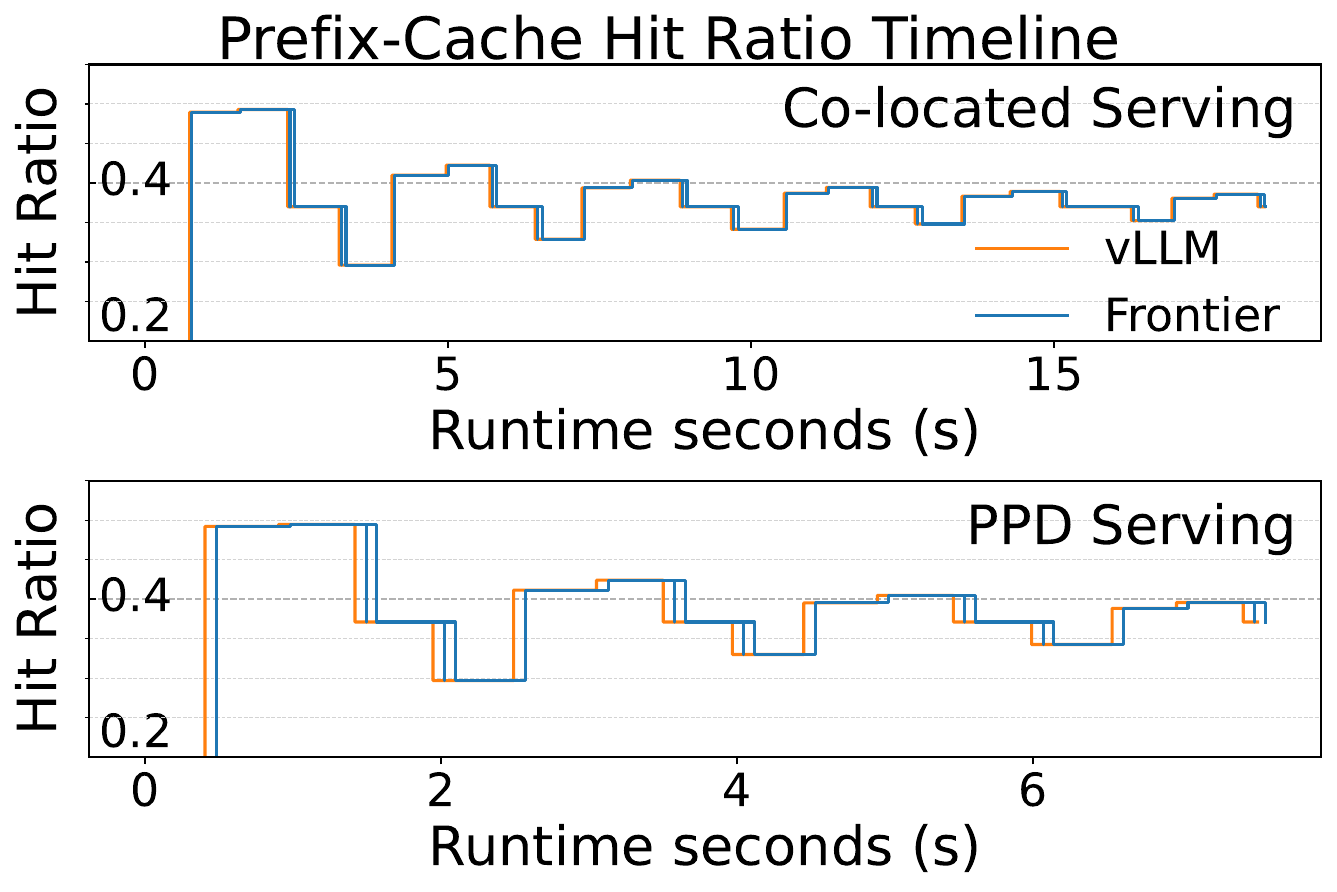}
        \vspace{-7mm}
        \caption{Cumulative prefix-cache hit ratio over time on Qwen3-30B MoE under co-location and PDD.}
        \label{fig:prefix_hit_ratio_timeline}
    \end{minipage}%
    \hfill%
    \begin{minipage}[t]{0.44\columnwidth}
        \centering
        \vspace{0pt}
        \vspace{2mm}
        \tiny
        \setlength{\tabcolsep}{1.2pt}
        \renewcommand{\arraystretch}{0.6}
        \resizebox{\linewidth}{!}{%
        \begin{tabular}{@{}cccccc@{}}
            \toprule
            VL & AR & TTFT (\%) & TPOT (\%) & Thpt. (\%) & E2E (\%) \\
            \midrule
            \multirow{2}{*}{2}  & 0.3 & 5.48 & 4.37 & 0.77 & 5.01 \\
                                & 0.7 & 2.75 & 1.92 & 1.88 & 2.40 \\
            \midrule
            \multirow{2}{*}{8}  & 0.3 & 0.84 & 0.00 & 5.12 & 6.45 \\
                                & 0.7 & 4.43 & 1.26 & 3.53 & 5.57 \\
            \midrule
            \multirow{2}{*}{32} & 0.3 & 4.14 & 6.18 & 4.23 & 9.32 \\
                                & 0.7 & 6.40 & 3.25 & 2.05 & 11.28 \\
            \bottomrule
        \end{tabular}}%
        \captionof{table}{Speculative decoding (MTP) fidelity on Qwen3-30B MoE under SharedGPT. Values are relative errors in percent. VL: verify length; AR: forced acceptance ratio.}
        \label{tab:specdecode_mtp_pass}
    \end{minipage}
    \vspace{-5mm}
\end{figure}

\begin{figure*}[t]
    \centering
    \includegraphics[width=0.9\linewidth]{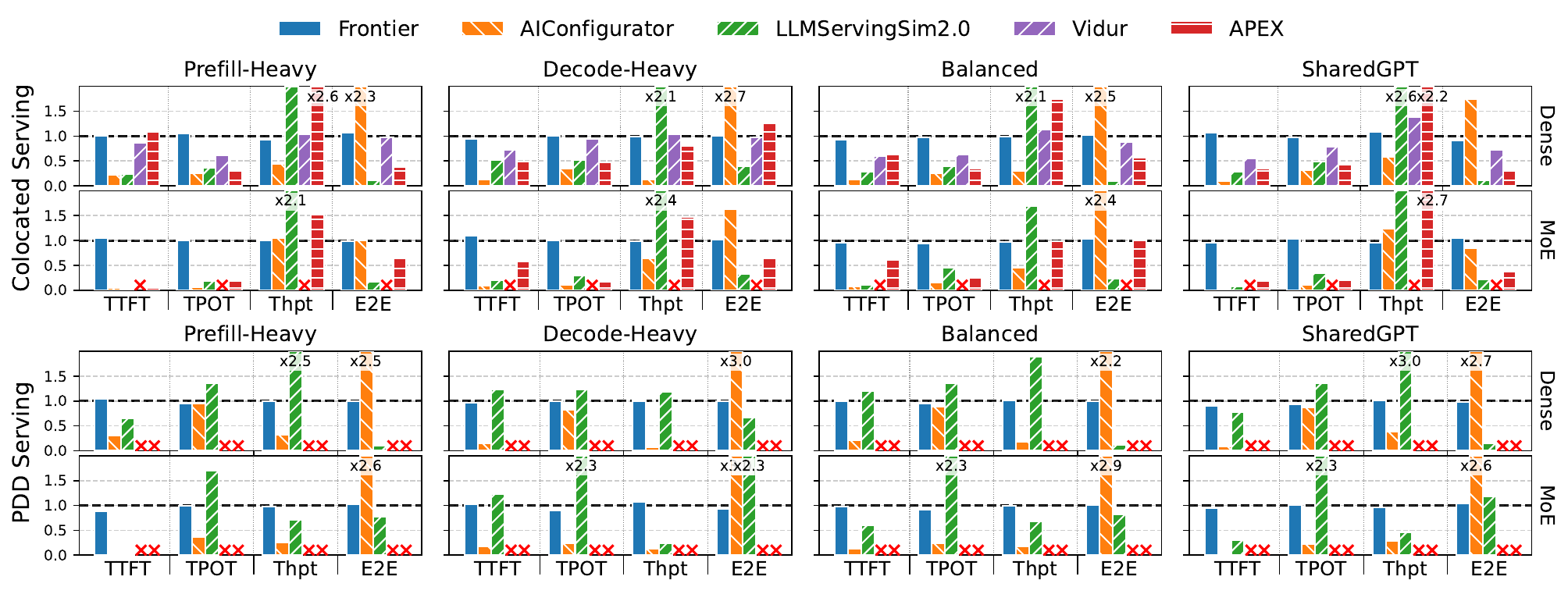}
    \vspace{-5mm}
    \caption{End-to-end fidelity on a 16-card H800 testbed for co-location and PDD across prefill-heavy, decode-heavy, balanced, and SharedGPT workloads. Panels report TTFT, TPOT, throughput, and E2E makespan for both Llama3.1-8B (dense) and Qwen3-30B MoE; dashed bars indicate simulators that do not support the serving architecture or model family. All data are normalized against the ground truth (vLLM), represented by the black line (y=1.0).}
    \label{fig:co_pdd}
\end{figure*}

\noindent\textbf{Memory (KV-cache budget) accuracy.}
Figure~\ref{fig:moe_colocation_sharedgpt_kv_budget_timeline} and Table~\ref{tab:initial_kv_budget_profiled} report the initial and time-varying per-rank KV-cache block budget on Qwen3-30B MoE under co-location and PDD.
\sys keeps the initial block budget within $1.89\%$ of vLLM across all four $(pp,tp,dp,ep)$ configurations, whereas the analytical baseline shared by most prior simulators over-reports the admissible budget by $14.10\%$--$39.73\%$; along the SharedGPT replay the block-availability curve of \sys overlaps the vLLM curve through admission, preemption, and release events, and the simulated makespan lies within $7.6\%$ of the measurement.
The gap stems from the KV-cache manager's treatment of other memory overhead: parameter memory is one of several occupants of the device budget, and omitting the runtime footprint of activation scratch, NCCL workspace, and CUDA Graph capture regions inflates the admissible block count that the scheduler later consumes for admission, chunking, and watermark-triggered preemption.
The \sys overhead residual is not uniform across configurations: it concentrates on pipeline-parallel layouts where a single representative-rank profile misses the PP-worst slice, so the direct tightening path is per-stage reprofiling rather than refining the analytical formula.

\subsection{Runtime Optimization Fidelity}
\label{subsec:runtime_opt_fidelity}

\noindent\textbf{CUDA Graph.}
Figure~\ref{fig:cuda_graph_sharedgpt_speedup} and Table~\ref{tab:cuda_graph_token_accounting} report decode throughput and compute-participating token counts for Qwen3-30B MoE under vLLM's \texttt{full\_decode\_only} capture path on SharedGPT.
\sys matches vLLM's decode speedup within $1.7\%$ under co-location and $6.1\%$ under PDD, preserving the $2.86$--$3.05\times$ envelope; its padded token count tracks vLLM within $2.55\%$ across prefill-heavy, decode-heavy, and SharedGPT workloads.
AIConfigurator, which omits graph-capture modeling, instead produces a workload-insensitive TPOT surface that structurally diverges from measurements (\S\ref{subsec:challenges}).
The residual reflects CUDA Graph's coupled effects: batch quantization inflates compute-participating tokens while graph replay removes launch overhead, so the Fidelity Plane must switch to kernel-only timings in sync with padding---a coupling no scalar speedup factor can reproduce.
The $6.1\%$ PDD residual comes from the decode-side padding tail under a sparse capture-bin ladder, suggesting denser capture-bin coverage rather than a new modeling primitive.

% \vspace{0.2em}

\noindent\textbf{Prefix cache.}
Figure~\ref{fig:prefix_hit_ratio_timeline} reports cumulative prefix hit ratios on Qwen3-30B MoE under co-location and PDD with vLLM's block-hash prefix cache enabled.
\sys matches vLLM's final hit ratios exactly: $36.98\%$ under co-location and $37.11\%$ under PDD.
The match comes from modeling prefix caching as a block-hash index that marks matched prefix blocks as already computed, thereby reshaping scheduler-visible admission state without storing KV tensors.
This replays vLLM's block-level hit/miss sequence and admission behavior while avoiding token-level KV-cache bookkeeping, which is orthogonal to admission-shape fidelity.

% \vspace{0.2em}
\noindent\textbf{Speculative decoding.}
Table~\ref{tab:specdecode_mtp_pass} reports p95 fidelity of \sys on Qwen3-30B MoE MTP under SharedGPT for verify-token lengths $\{2,8,32\}$ and forced acceptance ratios $\{0.3,0.7\}$.
Across the six configurations, \sys keeps TTFT, TPOT, throughput, and E2E p95 errors within $11.28\%$, with five below $10\%$.
% The worst case, $11.28\%$ E2E error at VL=$32$ and AR=$0.7$.
% AIConfigurator's analytical MTP path instead incurs $27.2\%$--$66.6\%$ decode-speedup error and flips the trend on low-acceptance batches.
Among existing simulators, only AIConfigurator provides limited support for MTP.
AIConfigurator's analytical MTP path instead incurs an average $46.9\%$ decode throughput error and flips the trend on low-acceptance (AR=0.3) batches.
This gap comes from event-level MTP modeling: \sys tracks per-request planned, verified, accepted, and committed tokens, and models MTP as a draft$\to$verify$\to$commit cycle sharing a batch with ordinary decode.
It therefore preserves per-request speculative-depth and acceptance variance within a batch, which governs tail latency but is lost in static analytical speedup models.
% The worst case, $11.28\%$ E2E error at VL=$32$ and AR=$0.7$, stems from wide-verify bookkeeping overhead outside current operator-timing coverage; extending profiling to wide-verify token buckets is the direct fix.

\vspace{-1em}
\subsection{End-to-End Fidelity}
\label{subsec:e2e_fidelity}

% \vspace{0.2em}
\noindent\textbf{Co-location.}
Under co-location (Figure~\ref{fig:co_pdd}, left), \sys tracks vLLM within $9.37\%$ across $32$ cases spanning Llama3.1-8B and Qwen3-30B MoE, with TPOT-p95 and E2E makespan inside $6.4\%$ in all but one case, whereas Vidur (dense-only) errors range from $2.9\%$ to $45.5\%$.
While effective for TensorRT-LLM on H200~\cite{xu2026aiconfiguratorlightningfastconfigurationoptimization}, AIConfigurator prediction accuracy degrades under vLLM and H800.
It produces E2E-makespan errors up to $170.0\%$ on decode-heavy dense and $135.3\%$ on balanced MoE.
Vidur's accuracy collapses on MoE because its operator library has no GroupedGEMM class and cannot express routing-dependent runtime, while AIConfigurator's decode-heavy makespan over-prediction is the compounded consequence of missing CUDA-graph and speculative paths (\S\ref{subsec:challenges}) once they are composed across hundreds of decode steps.
\sys remains within this envelope because the event-driven \textit{execution plane} keeps the closed loop of per-iteration cost, KV-cache block capacity, and admission-watermark preemption intact; small per-operator errors bounded in \S\ref{subsec:op_fidelity} therefore do not amplify through the feedback path into the qualitatively different regimes that static models exhibit.
% The residual clusters on prefill-heavy SharedGPT throughput ($9.03\%$ dense) and decode-heavy TTFT ($9.37\%$ MoE), both traceable to tail variance at admission that the current scheduler mirrors from vLLM but not yet with matching tail statistics.

\vspace{0.2em}
\noindent\textbf{PDD.}
Under PDD (Figure~\ref{fig:co_pdd}, right; $tp{=}4,dp{=}2$ for dense and $tp{=}4,dp{=}2,ep{=}8$ for MoE), \sys keeps all $32$ metric entries within $10.99\%$ of vLLM, with $29$ of $32$ inside $9\%$ and $20$ of $32$ inside $3\%$; Vidur and Apex do not support PDD and therefore cannot be placed on this axis, while AIConfigurator reaches $200.0\%$ E2E-makespan error on both dense and MoE decode-heavy.
The AIConfigurator collapse is architectural: its homogeneous-replica model cannot represent the cross-cluster KV-cache transfer dependency, so a missing arc in the event graph translates into a multiplicative makespan error that no post-hoc scalar can close (\S\ref{subsec:challenges}).
\sys holds because the \textit{control plane} materializes PDD as explicit cross-cluster KV-cache transfer events, and the \textit{fidelity plane} resolves each transfer against KV-cache size, link topology, and transfer concurrency (\S\ref{subsec:control_plane}), preserving the strict prefill$\to$transfer$\to$decode ordering that governs queueing.
We found the worst case is MoE prefill-heavy TTFT at $10.99\%$, which falls in the regime where chunked-prefill residuals interact with expert-routing imbalance at admission; tightening it requires extending the wide-chunk token buckets in the operator profile rather than new control-plane primitives.

\begin{figure}[t]
    \centering
    \includegraphics[width=0.85\linewidth]{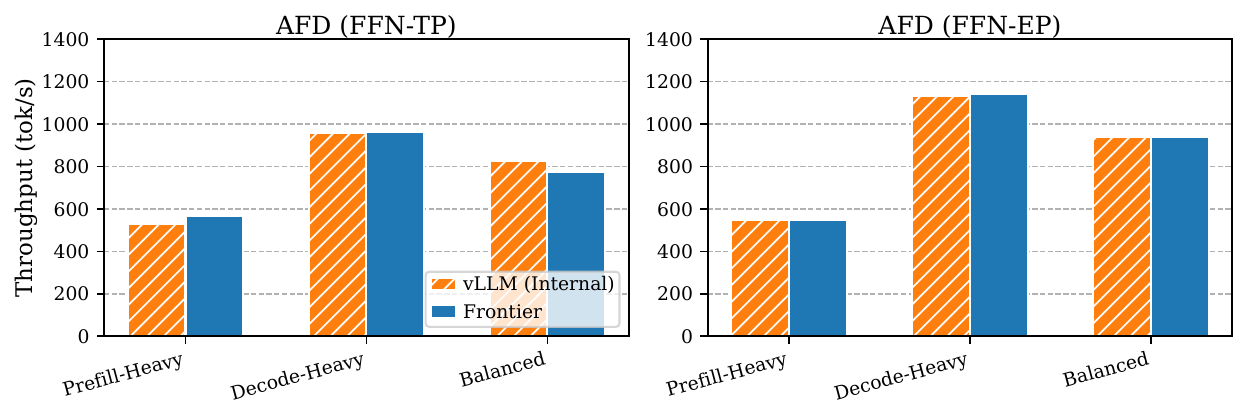}
    \vspace{-5mm}
    \caption{AFD fidelity on Step3-316B (16 H800 GPUs) against the ground truth across prefill-heavy, decode-heavy, and balanced workloads. AFD-TP and AFD-EP report throughput (decode toks/s).}
    \label{fig:afd}
    \vspace{-2mm}
\end{figure}

\vspace{0.2em}
\noindent\textbf{AFD.}
We evaluate two AFD decode-FFN layouts on Step3-316B with decode-attention fixed at $dp{=}8$: FFN-TP shards expert weights with $tp{=}8$ and no EP, while FFN-EP enables expert parallelism with $ep{=}8$.
On both AFD-TP and AFD-EP (Figure~\ref{fig:afd}), \sys holds TPOT within $6.4\%$ and throughput within $7.0\%$ across prefill-heavy, decode-heavy, and balanced workloads, and reproduces decode-heavy TPOT to within $1.0\%$ on AFD-EP.
Architectural fidelity follows from the cluster-role decomposition of \S\ref{subsec:control_plane}: AFD is realized as a prefill cluster plus separate attention and FFN clusters with a paired attention-to-FFN transfer per decode iteration, and every MoE layer gates its combine collective on the slowest rank so routing-induced stragglers arise from per-rank prediction rather than a separate imbalance model.
Because AFD was calibrated in an offline setting, we focus the main comparison on throughput-oriented metrics, where decode iteration cost and cross-cluster transfer dominate sustained serving behavior.

\begin{figure}[t]
    \centering
    \includegraphics[width=\linewidth]{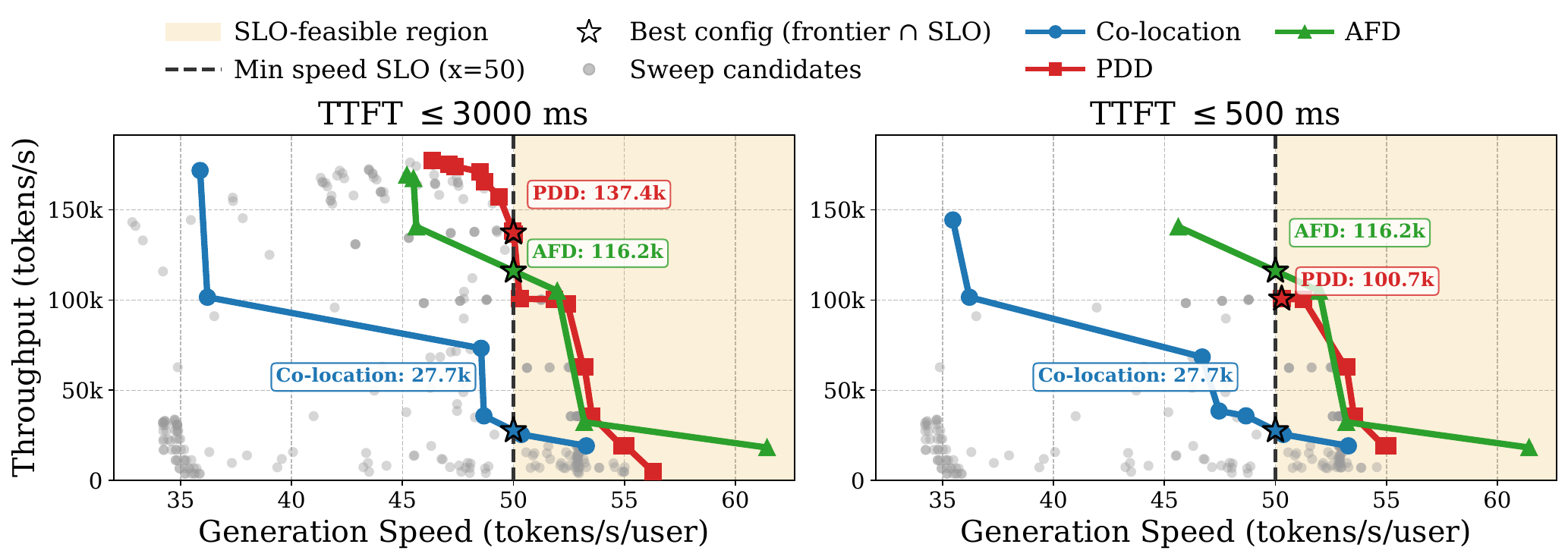}
    \vspace{-7mm}
    \caption{\sys projects throughput--generation-speed Pareto frontiers
    across co-location, PDD, and AFD on $256{\times}$H800 GPUs. The dashed
    line marks a $50$\,toks/s/user SLA; stars mark the best frontier point
    for each architecture under that SLA.}
    \Description{Two-panel Pareto frontier plots comparing co-location,
    PDD, and AFD under TTFT thresholds of 3000\,ms and 500\,ms.
    The x-axis is generation speed, inverse p95 TPOT, the y-axis is
    throughput, the shaded region satisfies the generation-speed SLA, and
    stars mark per-architecture best frontier points.}
    \label{fig:use_case_pareto_frontier}
    \vspace{-2mm}
\end{figure}

%!TEX root = newmain.tex
\section{Use Cases}
\label{sec:use_cases}

\noindent
We demonstrate \sys through four use cases in two groups.
The first group (\S\ref{subsec:use_case_pareto_frontier},
\S\ref{subsec:use_case_heterogeneous_disaggregated}) is
\emph{resource configuration tuning} for non-reasoning batch serving:
tuning the serving architecture and parallelism configuration for a given cluster (\S\ref{subsec:use_case_pareto_frontier}), and tuning the
GPU placement across heterogeneous GPU types
(\S\ref{subsec:use_case_heterogeneous_disaggregated}).
The second group (\S\ref{subsec:use_case_agentic_multiphase},
\S\ref{subsec:use_case_agentic_rl_rollout}) is \emph{what-if validation
of new optimizations} under emerging reasoning and RL rollout workloads:
validating a new scheduling algorithm
(\S\ref{subsec:use_case_agentic_multiphase}), and validating a dynamic
parallelism reconfiguration strategy
(\S\ref{subsec:use_case_agentic_rl_rollout}).
To the best of our knowledge, \emph{none of the four cases can be
reproduced as configured by any prior simulator}.

% \vspace{-1em}
\subsection{Optimal Configuration via Pareto Frontiers}
\label{subsec:use_case_pareto_frontier}

\noindent\textbf{Problem.}
Given fixed cluster resources, which serving architecture and parallelism
configuration is optimal under a given SLA constraint?

\noindent\textbf{Scenario.}
Pareto frontiers are the operator interface for architecture selection:
given a TTFT constraint, they show the maximum throughput reachable at each
generation-speed target.  Measuring this surface on a live $256$-GPU
cluster is impractical, since each point can change the parallelism
layout, cluster split, batching limit, memory budget, and load level.
We run the search ex-situ for a Llama-3.3-70B~\cite{Llama-3.3-70B-Instruct} batch-inference
workload on 256 H800 GPUs, comparing co-location, PDD, and AFD with
vLLM-style scheduler, CUDA Graph, and chunked prefill.  Across the three
serving architectures, the full sweep contains $483{,}536$ candidate
configurations under the 256-GPU budget.  Static memory filtering skips
$65{,}190$ OOM-infeasible cases before execution.
The final figures summarize 496 cases that satisfy the SLA.

\noindent\textbf{Simulation.}
Figure~\ref{fig:use_case_pareto_frontier} asks which serving architecture
delivers the most throughput once generation speed must exceed
$50$\,toks/s/user; the shaded region is feasible and the stars mark
the best frontier point at this SLA.  The three curves expose different
failure modes.  Co-location keeps all work in one scheduler and one
KV-cache budget, so prefill and decode compete for batch slots; once
p95 TPOT must stay near $20$\,ms (i.e., $50$\,toks/s/user), decode batches shrink and throughput
is capped.  PDD removes that interference by giving prefill and decode
separate clusters, which makes it strong when TTFT is loose and most
GPUs can be kept on decode.  AFD further separates decode-attention from
decode-FFN; it pays activation transfer between the two decode clusters,
but isolates the latency-sensitive attention path from FFN batching and
keeps the prefill path short.

The best-config markers quantify this tradeoff.  Under the loose
TTFT constraint (TTFT $\le 3000$\,ms), PDD is best: the SLA frontier point
reaches $137.4$K toks/s on the $\mathrm{P{:}D}=112{:}144$ split, where
more GPUs are assigned to decode.  AFD reaches $116.2$K toks/s with
$\mathrm{P{:}DA{:}DF}=120{:}80{:}56$, while co-location reaches only
$27.7$K toks/s.  When TTFT tightens to $500$\,ms, the optimum shifts:
AFD remains at $116.2$K toks/s with p95 TTFT $115.9$\,ms, whereas PDD
drops to $100.7$K toks/s at $\mathrm{P{:}D}=144{:}112$ and p95 TTFT
$211$\,ms because the prefill pool must grow at the expense of decode
capacity.  Without \sys that jointly models disaggregation,
parallelism, runtime state, and SLA filtering, this decision would be
made on the wrong frontier: PDD would appear universally preferable
under the loose-SLA view, but AFD is the correct choice under sub-second
TTFT.

\subsection{Heterogeneous GPUs for Disaggregated Serving}
\label{subsec:use_case_heterogeneous_disaggregated}

\begin{figure}[t]
  \centering
  \includegraphics[width=1\linewidth]{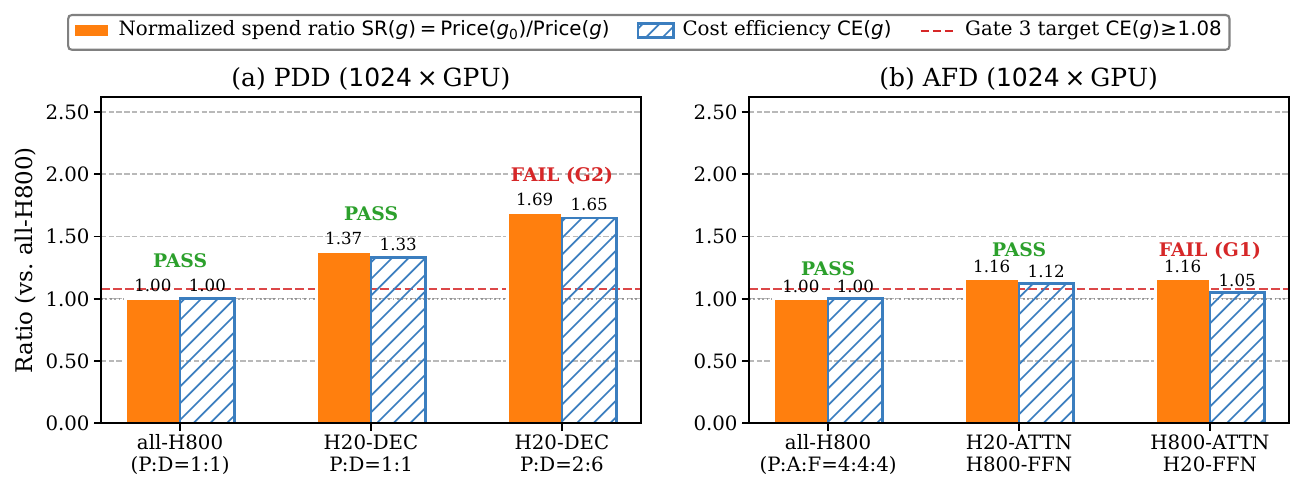}
  \vspace{-7mm}
  \caption{Heterogeneous Qwen3-235B-A22B~\cite{Qwen3-235B-A22B} allocation exposes which PDD and AFD role assignments convert hardware discounts into cost efficiency.}
  \label{fig:hetero_pd_pdaf_cost}
  \vspace{-3mm}
\end{figure}

\noindent\textbf{Problem.}
% Given heterogeneous GPUs, which role-to-GPU-type assignment maximizes cost efficiency while meeting SLA constraints? \hx{revise}
In a disaggregated serving architecture, how does the use of heterogeneous GPUs affect the performance-cost tradeoff?

\noindent\textbf{Scenario.}
% Heterogeneous GPU fleets are now a cost lever for large-scale serving:
% compute-dense devices and memory-bandwidth-oriented devices coexist, and
% operators want to match each serving role to the cheaper GPU type that fits
% its profile. 
It is known that PDD and AFD can improve serving efficiency. Here we explore another degree of freedom which is common in practice: heterogeneous GPU fleets. 
% For example, a cluster may contain both H800 and H20 GPUs, which have different price points and performance characteristics. 
We consider a cluster of fixed scale with 1024 GPUs, serving Qwen3-235B-A22B MoE~\cite{Qwen3-235B-A22B} along with the runtime optimizations specified in \cref{subsec:use_case_pareto_frontier}.
By assigning different GPUs to different roles (P/D/A/F), how does this further improve the cost-performance tradeoff?
% consistent runtime features.
% The assignment surface is defined by the disaggregated
% serving architecture: PDD exposes two roles, while AFD further splits
% the decode phase into attention and FFN roles, lifting placement to a
% three-way choice. 

\noindent\textbf{Simulation.}
Let $\mathrm{Price}(g){=}\sum_{r}\!N_r\,p(g(r))$ denote the cluster's total
hourly spend under an allocation scheme $g$, where $N_r$ is the number of GPUs
assigned to role $r$ and $p(\cdot)$ maps each GPU type to its public cloud
price (H800: \$$3.49$/hr, H20: \$$1.59$/hr)~\cite{cloudgpu}. 
Against the all-H800 baseline $g_0$, we report two scores for a candidate allocation:
\begin{equation*}
\mathrm{SR}(g)=\frac{\mathrm{Price}(g_0)}{\mathrm{Price}(g)},\quad
\mathrm{CE}(g)=\frac{T(g)\,/\,\mathrm{Price}(g)}{T(g_0)\,/\,\mathrm{Price}(g_0)},
\end{equation*}
where $T$ is aggregated token throughput.
$\mathrm{SR}(g)$ is the normalized spend ratio: higher is better, with
$\mathrm{SR}(g){>}1$ meaning $g$ is cheaper than $g_0$ on the bill.
$\mathrm{CE}(g)$ is cost efficiency, i.e. throughput-per-dollar relative to $g_0$: higher is
better, with $\mathrm{CE}(g){>}1$ meaning $g$ delivers more tokens per
dollar than $g_0$.  
Each valid allocation scheme must satisfy three constraints: 
(1) {\emph{hardware--workload alignment}: \sys uses per-role stage metrics and matched GPU-type counterfactuals to classify each role's bottleneck under the current workload; compute-bound roles must use H800, while H20 is allowed only when the role is not compute-limited.}
(2) {SLA constraints}: P95 TTFT and TPOT shall stay within the SLA thresholds; and 
% (3) {positive efficiency gain}, where $\mathrm{CE}(g)$ meaningfully exceeds $1$ while the P95 latency regression stays bounded \hx{what regression? same as SLA, no?}. 
% We set the target as
% $\mathrm{CE}(g){>}1.08$, an 8\% margin above break-even that leaves room
% for measurement variance and operational noise.
{(3) positive throughput-per-dollar ROI: even when SLA is met, a placement
that merely saves cost while noticeably reducing total throughput is not
deployment-worthy.  We require $\mathrm{CE}(g){>}1.08$, i.e., at least
8\% more tokens per dollar than the all-H800 baseline, absorbing
measurement variance and ensuring genuine operational value.}

These representative PDD and AFD placements lead to two findings
(Figure~\ref{fig:hetero_pd_pdaf_cost}).
First, \emph{heterogeneity does
not always pay off}.  Pushing the P:D capacity split toward decode in
PDD amplifies the H20 hourly discount: $\mathrm{SR}(g)$ climbs to
$1.69$ at $\text{P:D}{=}2{:}6$ and the nominal $\mathrm{CE}(g)$ reaches
$1.65$ as the expanded decode pool boosts token generation throughput.
However, P capacity collapses, a long queueing tail
dominates P95 TTFT, and the candidate is rejected at Gate~2; the price
saving is invalidated by an SLA-violating queueing tail.  Second,
\emph{heterogeneous savings are gated by different bottlenecks in the
two disaggregated designs}.  PDD reaches its accepted point by
downgrading the whole D role at $\text{P:D}{=}1{:}1$
($\mathrm{SR}(g){=}1.37$, $\mathrm{CE}(g){=}1.33$); pushing the split
toward decode keeps amplifying the nominal discount but starves
P, so the candidate is rejected by queue-driven SLA
violation.  AFD, in contrast, fixes $\text{P:A:F}{=}4{:}4{:}4$ and asks
which decode sub-role to downgrade: placing H20 on A
passes at $\mathrm{SR}(g){=}1.16$, $\mathrm{CE}(g){=}1.12$, while the
symmetric swap (H20 on F) is rejected at Gate~1 because
the FFN role is compute-bound and regresses on H20
($\mathrm{CE}(g){=}1.05$).  A price-only or fused-decode analysis would
favor rejected placements, either because they maximize bill savings or
because they cannot express attention/FFN role-GPU-type mismatch.
\sys's disaggregated cost-performance modeling ties GPU-type choice to role
bottlenecks and SLA outcomes, separating raw price reduction from
deployment-grade cost efficiency.

% \vspace{-0.5em}
\subsection{Scheduling Algorithm Validation for Stateful Reasoning}
\label{subsec:use_case_agentic_multiphase}

\begin{figure}[t]
    \centering
    \includegraphics[width=1\linewidth]{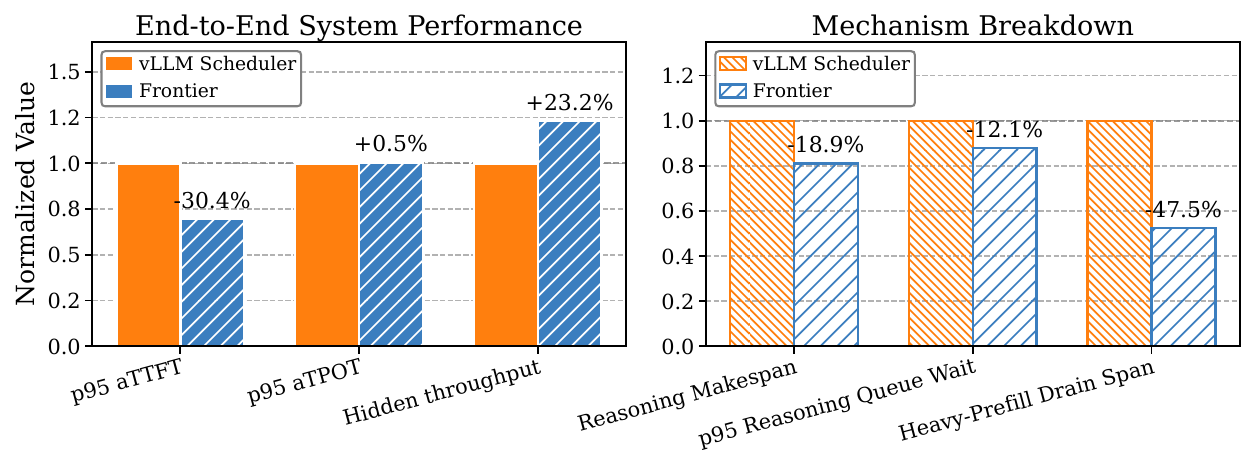}
    \vspace{-8mm}
    \caption{Phase-aware scheduling for multi-round reasoning.}
    \label{fig:agentic_multiphase}
    \vspace{-3mm}
\end{figure}

\noindent\textbf{Problem.}
Given a fixed resource configuration, can a new scheduling algorithm
that exploits cross-round request state improve latency over existing
priority policies?

\noindent\textbf{Scenario.}
Multi-round agentic reasoning workloads no longer behave like a single prompt
followed by a single decode stream.  Taking a coding agent such as
Claude Code~\cite{claudecode} as an example, each request progresses through two
phases: \emph{internal planning} and \emph{final response generation}.
During planning, the agent runs multiple thinking rounds marked by
\texttt{<thinking>} tags~\cite{guo2025deepseek}; each round may issue tool calls such as code
execution or calculation, whose observations seed the next round until
planning terminates and the visible response is produced.  The resulting
traffic is long-context and long-tailed~\cite{tan2026orchestrrl}: the number of thinking
rounds varies sharply across sessions, and per-round prefill spans a
wide range.  The essential modeling requirement is therefore not merely
longer contexts, but request-state and event modeling: the simulator
must preserve round history, tool-call delay, requeue events, and
prefix-cache continuity as part of the same session.  This changes the
scheduling problem from prioritizing isolated requests to \emph{scheduling
stateful, multi-event sessions} whose final answer latency depends on
hidden planning history.  Since short-job-first scheduling is effective
in non-reasoning serving---for example, FastServe~\cite{wu2023fast} uses a
skip-join multi-level feedback queue (MLFQ)~\cite{corbato1962experimental} to prioritize short requests---we ask whether such
mechanisms remain effective once the reasoning state is made explicit.
We use \emph{aTTFT} to denote answer-visible TTFT, measured from
session arrival to completion of the final answer-round prefill, i.e.,
the first user-visible token boundary.  We define hidden planning
throughput as the hidden planning-token rate over the fixed trace.

\noindent\textbf{Simulation.}
We employ \sys for an algorithmic exploration.
We model Llama-3.1-405B-FP8~\cite{Llama-3.1-405B-FP8} on $1024{\times}$H800 GPUs under a PDD architecture,
with chunked prefill, prefix caching, CUDA Graph, quantization (FP8) and CPU offloading
enabled to reflect a realistic deployment.  The trace is prefill-heavy
and long-tailed as the scenario describes:
per-round prefill spans 4K--32K tokens while per-round decode stays at $\sim$0.2K.
Plugging FastServe's skip-join MLFQ into
this setting yields only a 1.45\% p95 aTTFT improvement and degrades
hidden planning throughput by 2.3\%.  The reason is that its priority is
decided from the current round's observable size and ignores
cross-round history; a heavy-tail session whose final answer round looks
small is therefore treated like an ordinary short request.  Guided by
this diagnosis, we deploy a phase-aware scheduler that classifies
sessions by both current-round size and accumulated planning history,
achieving a 30.4\% p95 aTTFT reduction and 23.2\% hidden-planning-throughput
improvement (Figure~\ref{fig:agentic_multiphase}).
We provide more details in the Appendix~(\cref{app:agentic-workload,app:agentic-first-sj2q,app:h2q-br}).
With \sys, we enable rapid, low-cost design and validation of scheduling
algorithms for advanced scenarios like stateful reasoning---without
deploying large-scale GPU clusters or modifying a production serving stack.

% \vspace{-0.5em}
\subsection{Dynamic Parallelism Reconfig. for RL Rollouts}
\label{subsec:use_case_agentic_rl_rollout}

\begin{figure}[t]
    \centering
    \includegraphics[width=1\linewidth]{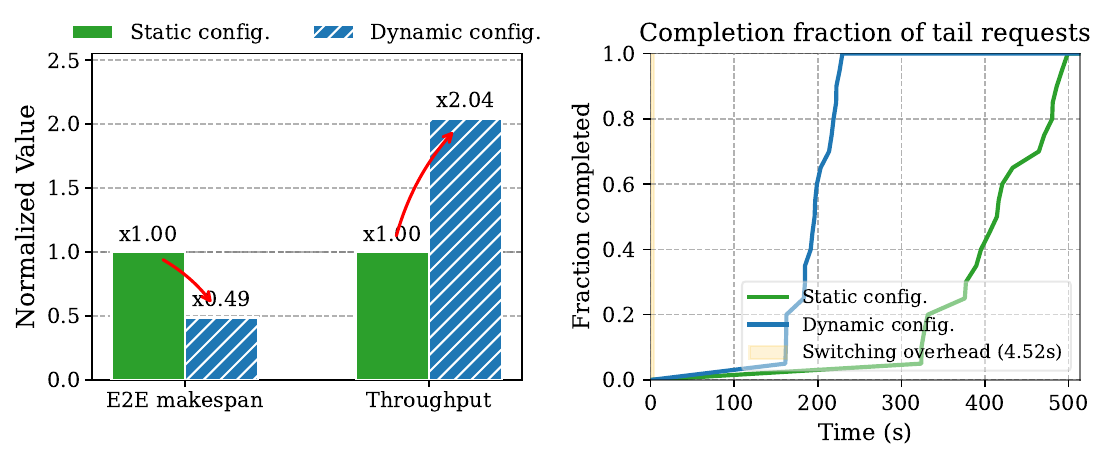}
    \vspace{-9mm}
    \caption{Performance gains from dynamic reconfiguration.}
    \label{fig:agentic_rl_rollout}
    \vspace{-3mm}
\end{figure}

\noindent\textbf{Problem.}
Given a fixed scheduling logic, can dynamically switching the parallelism
layout mid-workload reduce end-to-end makespan?

\noindent\textbf{Scenario.}
Agentic RL rollouts exhibit a phase transition that is invisible to
steady-state serving benchmarks~\cite{tan2026orchestrrl,zhang2026heddle}.  At the start of a rollout
batch, the trainer launches many trajectories in parallel, so the
inference cluster is bound by burst throughput and rewards high replica
concurrency.  Once most short trajectories terminate, a heavy-tail
subset continues decoding alone and dictates when the trainer can
advance to the next gradient step.  Recent work addresses this
asymmetry by reconfiguring parallelism online as the active set
shrinks~\cite{tan2026orchestrrl}.  The benefit depends on both
when to switch and how much reconfiguration cost the switch incurs.
We ask whether \sys can prototype such reconfiguration policies
and expose their system-level impact ahead of deployment.

\noindent\textbf{Simulation.}
We employ \sys for a dynamic reconfiguration exploration.
We retain the model, cluster scale, and runtime features from the
previous use case (\cref{subsec:use_case_agentic_multiphase}), and drive a co-located deployment with a
$4{,}000$-trajectory burst whose heavy-tail fraction is $5\%$.  The policy
is realized in \sys as a time-sliced layout switch.  The baseline pins
\emph{Layout~A}---a high-DP configuration
($\text{DP}{=}32,\text{PP}{=}16,\text{TP}{=}2$) that maximizes batch
parallelism---throughout the rollout.  The dynamic policy begins with
Layout~A and switches to \emph{Layout~B}
($\text{DP}{=}8,\text{PP}{=}16,\text{TP}{=}8$) once the
active-trajectory count drops below $10\%$ of the batch, paying a
profiled reconfiguration cost on transition, including weight reshard
and KV-cache rematerialization.  As
Figure~\ref{fig:agentic_rl_rollout} shows, the dynamic policy reduces
rollout makespan from $528.8\,\text{s}$ to $259.1\,\text{s}$ and improves
effective decode throughput by $2.04\times$.  The gain is
phase-matched parallelism: Layout~A absorbs the dense burst, while
re-sharding from replica-wide DP to wider TP after the tail emerges
shortens per-trajectory decode latency once spare replicas no longer
translate into utilization.  With \sys, such phase-adaptive policies can
be rapidly prototyped and validated before deployment, avoiding the
hardware cost of repeatedly sweeping thousand-GPU layouts and the
engineering cost of implementing layout switching directly in a
production serving stack.
% \vspace{-1em}

%!TEX root = newmain.tex
\section{Discussion}
\label{sec:discussion}

Currently, \sys is primarily developed and calibrated around vLLM. As outlined in \cref{sec:design}, its design natively supports extending simulation capabilities to other mainstream frameworks (e.g., SGLang~\cite{zheng2024sglang} and TensorRT-LLM~\cite{TensorRT}), which we leave for future work. For precision calibration, since GPU execution dominates the overall latency ($\sim$90\%)~\cite{agrawal2026revatitransparentgpufreetimewarp}, we follow Vidur's methodology to model the remaining CPU overhead by combining testbed measurements with a prediction mechanism (Fidelity Plane). However, this approach for CPU overhead prediction may still lack robustness.
As CPU overhead fluctuates with system scales and architectural updates, a potentially more accurate approach for CPU simulation would employ CUDA API interception—similar to Revati~\cite{agrawal2026revatitransparentgpufreetimewarp}—to enable full CPU-side replay. 
Nevertheless, the usability and scalability of this approach across diverse hardware, CUDA versions, and dynamic runtime optimizations (e.g., CUDA Graphs) remain debatable. 
Developing a robust yet accurate simulation scheme capable of adapting to diverse inference frameworks and technological advancements represents a promising future direction.
Furthermore, during calibration, we observed that beyond standard warmup, profiling tools easily introduce intrusive overhead. Therefore, we strictly isolate the measurement of fine-grained metrics (e.g., operator execution, batch scheduling) from system-level metrics (e.g., TTFT, TPOT, throughput) to prevent statistical contamination and ensure reliable calibration.
%!TEX root = newmain.tex
% \vspace{-0.3em}
\section{Related Work}
\label{sec:related}

\noindent We discuss related work other than those covered in \cref{sec:motivation}.

\noindent\textbf{GPU simulation.}
GPU computation simulators like SCALE-sim~\cite{samajdar2018scale} and Accel-Sim~\cite{khairy2020accel} focus on instruction-level modeling.
\sys can integrate these into the op library to support operation modeling for future hardware.

\noindent\textbf{Training simulators.}
Simulators for LLM training~\cite{feng2024echo,duan2024proteus,won2023astra} model training with a focus on optimizing parallelism strategies and resource utilization. 
However, they do not capture the dynamics of inference workloads and complex runtime behavior, which are critical for LLM serving. 

% \vspace{-1em}

% \noindent\textbf{Network simulation.} 
% % Tools like ASTRA-sim~\cite{won2023astra}, FlexFlow~\cite{jia2019beyond}, Proteus~\cite{duan2024proteus}, and NCCL-Predictor~\cite{nccl} rely on the ${\alpha-\beta}$ model~\cite{valiant1990bridging}, offering simplicity but often limited accuracy. 
% Discrete event simulators like ns-3~\cite{riley2010ns} and OMNeT++~\cite{varga2019practical} are well-known to have high overheads due to their packet-level whole-stack simulation. Other than parallelization optimizations \cite{GCLL23,BZTW24}, recently learning based approach to predict packet-level or flow-level performance without simulating the whole stack~\cite{ZNKY21,li2024m3,YPCL22} has shown promising results.
% As mentioned before, integrating them with \sys for training simulation is a promising direction for future work. 

%!TEX root = newmain.tex
% \vspace{-0.2em}

\section{Conclusion}
We present \sys, a discrete-event simulator for LLM serving.
By jointly modeling architectural heterogeneity, runtime behavior, and workload statefulness, \sys enables accurate, scalable evaluation of complex serving designs.
Results show that \sys reliably predicts performance across diverse configurations and supports large-scale design-space exploration on commodity hardware, providing a practical foundation for studying modern LLM serving systems.

\bibliographystyle{ACM-Reference-Format}
\bibliography{reference}

\clearpage
\appendix
%!TEX root = newmain.tex

\section{Additional Details}
\label{app:additional-details}

\subsection{H20 Operator Fidelity}
\label{app:h20-op-fidelity}

\begin{figure}[tbp]
  \centering
  \includegraphics[width=\columnwidth]{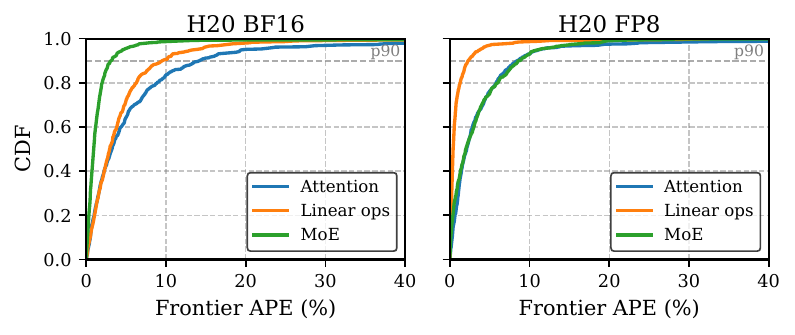}
  \caption{H20 operator fidelity CDFs. Curves show \sys absolute percentage error for attention, linear ops, and MoE on the H20 BF16 and FP8.}
  \label{fig:app-h20-op-fidelity}
\end{figure}

We use the selected CDF rows from the H20 BF16 and FP8 operator packages,
with 600 rows per family. Figure~\ref{fig:app-h20-op-fidelity} reports
\sys APE CDFs for attention, linear ops, and MoE of H20 BF16 and FP8 operator.

On BF16, \sys p50/p90 APE is 3.24\%/14.20\% for attention,
2.95\%/9.58\% for linear ops, and 0.93\%/3.12\% for MoE. On FP8, the
corresponding p50/p90 values are 2.11\%/8.49\%, 0.42\%/2.29\%, and
2.01\%/8.64\%. Linear ops stay the tightest curve in both packages.
Attention keeps the longest BF16 tail, while MoE widens again on FP8.

\subsection{H20 End-to-End Fidelity}
\label{app:h20-e2e-fidelity}

\begin{figure}[tbp]
  \centering
  \includegraphics[width=\columnwidth]{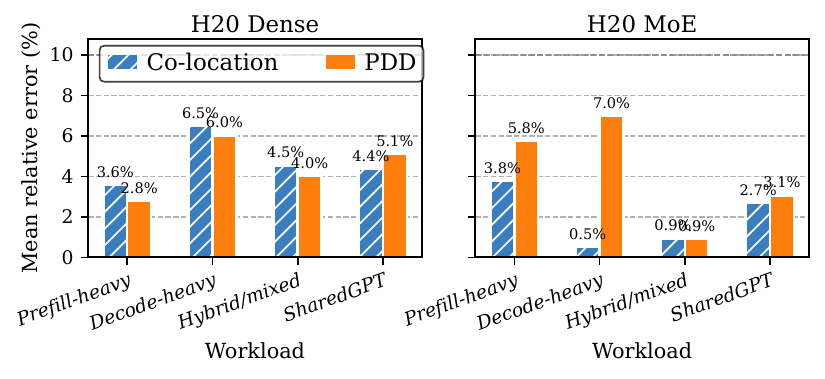}
  \caption{H20 end-to-end fidelity across dense and MoE model families. Bars show the mean relative error for co-location and PDD on the four workloads, averaging p95 TTFT, p95 TPOT, throughput, and p95 E2E within each workload-architecture pair.}
  \label{fig:app-h20-e2e-fidelity}
\end{figure}

Figure~\ref{fig:app-h20-e2e-fidelity} summarizes H20 end-to-end fidelity for dense and MoE models under co-location and PDD across prefill-heavy, decode-heavy, hybrid/mixed, and SharedGPT workloads. Each bar averages the relative error over p95 TTFT, p95 TPOT, throughput, and p95 E2E for one workload-architecture pair. All bars stay below 10\%; dense co-location peaks at 6.5\% on decode-heavy, dense PDD at 6.0\% on decode-heavy, MoE co-location at 3.8\% on prefill-heavy, and MoE PDD at 7.0\% on decode-heavy.

\begin{figure}[tbp]
  \centering
  \includegraphics[width=\columnwidth]{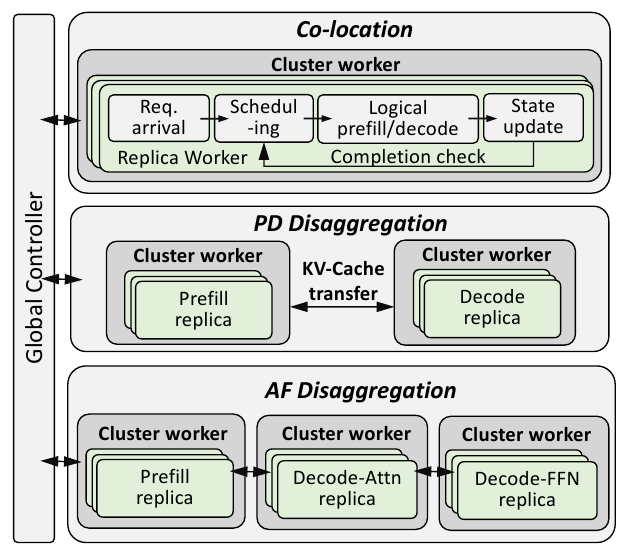}
  \caption{Example of three serving architectures in \sys.}
  \label{fig:three-arch}
  \vspace{-2.0em}
\end{figure}

\subsection{PDD and AFD Scheduling Workflow}
\label{app:pdd-afd-workflow}

\sys operates as a discrete-event simulator (DES). This subsection expands the
control and execution-plane mechanisms into a concrete workflow-level
algorithmic description.
We show the example of three serving architectures of \sys in Figure~\ref{fig:three-arch}.
Algorithm~\ref{alg:app-pdd-afd-workflow} illustrates
the scheduling and execution pipelines for both Prefill-Decode Disaggregation (PDD)
and Attention-FFN Disaggregation (AFD) serving architectures, using an MoE model
simulation with thinking/reasoning modes enabled as a representative example.

From a high-level perspective, the algorithm is modularly structured around role-local
scheduler ticks. In the initialization phase, roles are compiled and bound to their respective
parallel domains and hardware resources ($\{\mathsf{P}, \mathsf{D}\}$ for PDD, or
$\{\mathsf{P}, \mathsf{A}, \mathsf{F}\}$ for AFD). At runtime, an event-driven loop
processes request admissions and batch formations via \texttt{runtime\_adapter()} gates.
These gates cleanly inject production optimizations (e.g., chunked prefill, speculative decoding)
into the scheduler-visible state without polluting the core simulation logic.

Requests then branch into role-specific execution paths:
\begin{itemize}[leftmargin=1.5em,itemsep=0pt,topsep=0.2em]
\item $\mathsf{P}$ (Prefill): Completes prompt processing and triggers a request-granular KV-cache transfer to downstream decode roles.
\item $\mathsf{D}$ (PDD Decode): Executes unified Attention and FFN layers sequentially, emitting a global batch completion event.
\item $\mathsf{A}$ and $\mathsf{F}$ (AFD Decode): Pipelines per-layer activation transfers ($\mathsf{A} \to \mathsf{F}$ and $\mathsf{F} \to \mathsf{A}$). The $\mathsf{F}$ role explicitly manages MoE routing, Expert Parallel (EP) dispatch, and synchronization barriers.
\end{itemize}

For multi-phase reasoning workloads, requests traverse this role graph for each thinking round, re-entering the prefill queue after a simulated tool delay until the final answer is generated.

\begin{algorithm}[tbp]
\caption{PDD/AFD scheduling and execution workflow}
\label{alg:app-pdd-afd-workflow}
\small
\begin{algorithmic}[1]
\State \textbf{Compile roles:} Instantiate $\{\mathsf{P},\mathsf{D}\}$ (PDD) or $\{\mathsf{P},\mathsf{A},\mathsf{F}\}$ (AFD); bind each $c \in \mathcal{C}$ to domain, $Q_c$, budget, and backend.
\State \textbf{Arrival/re-entry:} Enqueue new requests and Thinking re-entries to $Q_{\mathsf{P}}$ at time $t$. \Comment{Preserve session affinity}
\For{each scheduler tick of role $c \in \mathcal{C}$}
    \State \textbf{Admission:} Call \texttt{runtime\_adapter()} on state. Admit requests, allocate blocks, and preempt while budget permits.
    \State \textbf{Execution:} Form batch $B$, call \texttt{runtime\_adapter()} on shape, query Fidelity Plane, and run through local PP stages.
    \If{role $c = \mathsf{P}$}
        \State \textbf{Prefill:} On completion, retain KV cache and emit request KV-cache transfer start. On end, enqueue $r$ to $\mathsf{D}$ or $\mathsf{A}$.
    \ElsIf{role $c = \mathsf{D}$ (PDD)}
        \State \textbf{PDD decode:} Execute unified Attention+FFN. Emit \texttt{GlobalBatchEnd} on final layer/token completion.
    \ElsIf{role $c = \mathsf{A}$ (AFD)}
        \State \textbf{AFD attention:} For non-final decode layers, emit A$\to$F transfer with $(\ell, \mathrm{stage}, \mathrm{lane})$ metadata.
    \ElsIf{role $c = \mathsf{F}$ (AFD)}
        \State \textbf{AFD FFN/MoE:} Group A$\to$F arrivals by stage/lane. Apply padding contracts, route tokens, build EP sub-batches.
        \If{$\mathrm{ep}_{\mathrm{ffn}} > 1$}
            \State Run pre-dispatch MoE work, wait for EP dispatch collective, then execute expert sub-batches.
            \State \textbf{EP sync:} \textbf{Wait} for all EP lanes; combine at $\max_i t^i_{\mathrm{ready}} + \Delta_{\mathrm{EP}}$.
        \Else
            \State Execute expert sub-batches locally without EP synchronization.
        \EndIf
        \State Recover raw batches and emit F$\to$A transfers. $\mathsf{A}$ advances layer progress or emits \texttt{GlobalBatchEnd}.
    \EndIf
    \State \textbf{Completion:} Commit tokens, release resources. Emit \texttt{ThinkingRequeue} if non-final, else export user metrics.
\EndFor
\end{algorithmic}
\end{algorithm}
\par\smallskip

The key property is that disaggregation changes the event graph, not the
simulator core.  PDD inserts a request-granular KV-cache dependency; AFD adds
per-layer activation ping-pong.  Both keep MoE synchronization explicit:
prefill and PDD decode use in-role sync barriers, while AFD exposes decode-time
FFN barriers in $\mathsf{F}$ before activations return to $\mathsf{A}$.
Runtime optimizations remain modular because adapter gates mutate
scheduler-visible state, batch shape, or per-request progress exactly where the
production loop observes those effects.

\makeatletter\global\@ACM@balancefalse\makeatother
\section{Agentic Multi-Phase Reasoning}
\label{app:agentic-multiphase}
\label{app:agentic}

\begin{table}[tbp]
\centering
\caption{Multi-phase reasoning workload.  Each cell is
$\ell_{r,r'}/o_{r,r'}$ (new prompt / decode tokens); round 5 is the
answer-visible round.}
\label{tab:app-agentic-workload}
\resizebox{\columnwidth}{!}{%
\begin{tabular}{lrrrrr}
\toprule
Template & $r{=}1$ & $r{=}2$ & $r{=}3$ & $r{=}4$ & $r{=}5$ \\
\midrule
Short      & 4096/96 & 1024/64 & 512/64 & 512/64 & 256/192 \\
Heavy tail & 32768/96 & 16384/64 & 8192/64 & 4096/64 & 256/192 \\
\bottomrule
\end{tabular}}
\end{table}

\subsection{Workload}
\label{app:agentic-workload}

\noindent\textbf{Multi-round requests.}
We model each agentic request $r$ as a sequence of $R_r{=}5$ rounds: four
hidden planning rounds (matching the \texttt{<thinking>} blocks in the
scenario description) followed by one answer-visible round.  Round $r'$
contributes $\ell_{r,r'}$ new prompt tokens (after same-request prefix
reuse) and $o_{r,r'}$ decode tokens.  Following the main definition,
\emph{aTTFT} is the time from request arrival to completion of the
answer-round prefill.  Table~\ref{tab:app-agentic-workload} lists the
two templates used in our trace.

\subsection{Why Skip-Join MLFQ Is Insufficient}
\label{app:agentic-first-sj2q}

\begin{figure}[tbp]
  \centering
  \includegraphics[width=0.7\columnwidth]{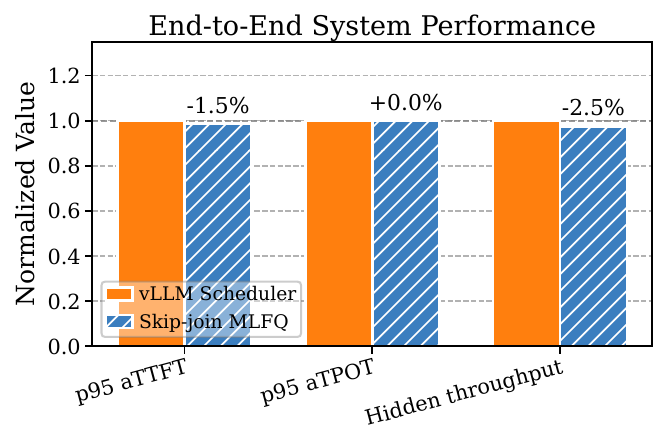}
  \caption{Skip-join MLFQ on the agentic trace: p95 aTTFT improves
  marginally and hidden planning throughput regresses. Blue-bar
  annotations show the relative change over vLLM.}
  \label{fig:app-first-sj2q}
\end{figure}

We use \sys to realize the skip-join MLFQ comparator against the native
vLLM scheduler. It yields only marginal gains: p95 aTTFT drops from
$20.84\,\text{s}$ to $20.54\,\text{s}$ ($-1.45\%$), p95 aTPOT is
unchanged, and hidden planning throughput \emph{regresses} by $2.54\%$
(Figure~\ref{fig:app-first-sj2q}). The blue-bar annotations in the figure
show the relative change over vLLM.

\noindent
The failure mode is structural: priority is decided on the
\emph{current} round's observable size, but a heavy-tail request may
present a small answer round indistinguishable from a true short request.
Symmetrically, a heavy hidden-prefill continuation can occupy the prefill
budget ahead of latency-critical answer prefills.  This motivates a
classifier that retains \emph{cross-round history}.

\subsection{H2Q-BR: History-Aware Two-Queue Scheduling with Bounded Release}
\label{app:h2q-br}

\noindent\textbf{Design intent.}
Skip-join MLFQ improves over FIFO-style admission by letting short
current-round work bypass long prompt chunks. However, the previous
section shows that this local view is not enough for agentic sessions:
a heavy-tail request can later expose a tiny answer-visible round, and a
running hidden-prefill continuation can keep consuming prefill budget
before short answer prefills are admitted. \textbf{H2Q-BR}
(\emph{History-Aware Two-Queue with Bounded Release}) keeps the
skip-join intuition, but adds session history so that the scheduler can
distinguish ``small now'' from ``historically small'' without using a
final-round oracle.

\noindent\textbf{High-level idea.}
H2Q-BR follows three simple rules. First, it records compact
\texttt{session\_id}-scoped history: whether a session has already
shown long-prompt behavior, how many new prompt tokens it has consumed,
the current round's new-prompt increment, and whether a prefill chunk
spilled across scheduling iterations. Second, it separates work into a
short queue $Q_S$ and a long-history queue $Q_L$. Small requests from
historically short sessions remain in $Q_S$ and are ordered by smaller
current prompts first; for example, a 256-token answer prefill from a
short session can bypass a 4096-token hidden prompt. A 32768-token
hidden prefill enters $Q_L$ immediately, and later slices from the same
session remain in $Q_L$ even if the current answer round is only
256 tokens. Third, H2Q-BR prevents starvation: long-history work normally
waits behind $Q_S$, but a one-shot carryover release lets an already
started long prefill make bounded progress, and a liveness quota forces
the oldest $Q_L$ slice after too many consecutive short-queue slices.

\noindent
The policy only changes request order before batch construction. The
inherited batch builder still enforces token budgets, memory admission,
chunking, and preemption; H2Q-BR does not replace the execution model.

\noindent\textbf{State update and classification.}
Algorithm~\ref{alg:app-h2q-br} starts when a slice arrives or re-enters
the scheduler. H2Q-BR computes the current-round prompt length as the
increase in total prefill tokens since the previous completed round,
then classifies the slice by Eq.~\ref{eq:h2q-classification}. Once a
session crosses the long-history boundary, the sticky flag keeps later
rounds in $Q_L$; there is no round-class oracle and no final-round
promotion.

\noindent\textbf{Classification.}
\begin{equation}
q_r =
\begin{cases}
Q_L, & z_r{=}1 \;\lor\; H_r>C \;\lor\; \ell_r>L,\\
Q_S, & \text{otherwise}.
\end{cases}
\label{eq:h2q-classification}
\end{equation}
Here $z_r$ is the sticky long-history flag, $H_r$ is cumulative served
new tokens, $C$ is the service cap, and $L$ is the long-round threshold.
The strict $\ell_r>L$ matters: a 4096-token short round stays in $Q_S$,
while a 32768-token heavy hidden round enters $Q_L$ on arrival.

\noindent\textbf{Bounded release.}
The middle procedure of Algorithm~\ref{alg:app-h2q-br} decides whether
any long-history slice should temporarily outrank $Q_S$. When chunked
prefill stops before finishing a prompt, H2Q-BR marks the session as
long-history, sets a one-shot carryover-release flag, and requeues the
slice in $Q_L$. On a later scheduling pass, it may release at most one
such carryover request. The release is safe only if the request arrived
no later than the oldest waiting $Q_S$ request; if no $Q_S$ request is
waiting, the oldest pending carryover request is released. This gives
spilled prefills a bounded path forward without letting them jump ahead
of older short requests.

\noindent\textbf{Ranking and batch construction.}
Before each scheduling pass, H2Q-BR ranks the combined running and
waiting sets by
\begin{equation}
\rho(i)=
\begin{cases}
-2, & i=i_{\mathrm{rel}}\quad\text{(release)},\\
-1, & i=i_{\mathrm{live}}\quad\text{(liveness)},\\
\;\;0,  & q_{r(i)}=Q_S,\\
\;\;1,  & q_{r(i)}=Q_L,
\end{cases}
\label{eq:h2q-rank}
\end{equation}
where $i_{\mathrm{live}}$ is the oldest $Q_L$ slice forced after the
short-streak counter reaches $B$. Within $Q_S$, the secondary key is
$(\ell_r,\,d_i,\,a_i)$: smaller prompts first, prefill before decode,
then earlier arrival. Within normal $Q_L$, decode precedes prefill to
limit TPOT regression. A released or liveness-forced slice uses the
negative rank above, then decode-before-prefill and arrival time as
tie-breakers. After this ordering step, the existing scheduler consumes
the ordered running and waiting sets exactly as before. The final
procedure in Algorithm~\ref{alg:app-h2q-br} then accounts executed
tokens, refreshes per-session history, consumes a selected carryover
credit, and updates the short-streak counter for the next pass.

\begin{algorithm}[tbp]
\caption{H2Q-BR scheduling step}
\label{alg:app-h2q-br}
\small
\begin{algorithmic}[1]
\Require Session state $(z_r, H_r, T_r^{\mathrm{last}}, c_r)$ for each session $r$. Short streak counter $\eta$.
\Statex
\Procedure{OnArrivalOrReentry}{slice $i$, session $r$}
    \State Load session state $(z_r,H_r,T_r^{\mathrm{last}},c_r)$ from \texttt{session\_id}.
    \State $\ell_r \gets \max(\textsc{PrefillTokens}(i) - T_r^{\mathrm{last}}, 0)$ \Comment{Current-round prompt length}
    \If{$z_r \lor (H_r > C) \lor (\ell_r > L)$}
        \State $q_r \gets Q_L$ \Comment{Assign to long-history queue}
    \Else
        \State $q_r \gets Q_S$ \Comment{Assign to short-history queue}
    \EndIf
    \State Enqueue slice $i$ to $q_r$.
\EndProcedure
\Statex
\Procedure{BeforeSchedulingPass}{}
    \State \textbf{Bounded Release:}
    \If{$Q_S = \emptyset$}
        \State $i_{\mathrm{rel}} \gets$ oldest pending carryover slice (with $c_{r(i)}=1$)
    \Else
        \State $i_{\mathrm{rel}} \gets$ oldest pending carryover slice (with $c_{r(i)}=1$) where $a_{i_{\mathrm{rel}}} \le \min_{j \in Q_S} a_j$
    \EndIf
    \If{$\eta \ge B$ \textbf{and} $Q_L \neq \emptyset$}
        \State $i_{\mathrm{live}} \gets \arg\min_{i \in Q_L} a_i$ \Comment{Force liveness for oldest $Q_L$}
    \Else
        \State $i_{\mathrm{live}} \gets \bot$
    \EndIf
    \State Rank \textsc{Running} $\cup$ \textsc{Waiting} by $\rho(i)$ (Eq.~\ref{eq:h2q-rank}).
    \State Apply tie-breakers: $(\ell_r, d_i, a_i)$ for $Q_S$, and $(d_i, a_i)$ for $Q_L$.
    \State Pass the ordered queues to the batch builder.
\EndProcedure
\Statex
\Procedure{OnBatchCompletion}{}
    \For{each scheduled slice $i$}
        \State $H_{r(i)} \gets H_{r(i)} + \text{executed new tokens}$.
        \If{round completes}
            \State Refresh $T_{r(i)}^{\mathrm{last}}$.
        \EndIf
        \If{prefill slice $i$ makes partial progress but is unfinished}
            \State $z_{r(i)} \gets 1$, $c_{r(i)} \gets 1$, $q_{r(i)} \gets Q_L$ \Comment{Mark carryover}
        \EndIf
        \If{$i = i_{\mathrm{rel}}$ \Comment{The selected release slice ran}}
            \State Clear its carryover flag $c_{r(i)} \gets 0$.
        \EndIf
    \EndFor
    \If{any $Q_L$ slice ran}
        \State $\eta \gets 0$
    \Else
        \State $\eta \gets \eta + (\text{number of scheduled } Q_S \text{ slices})$
    \EndIf
\EndProcedure
\end{algorithmic}
\end{algorithm}
\par\smallskip

% \subsection{Mechanism Evidence}
% \label{app:h2q-evidence}

% The right panel of Figure~\ref{fig:agentic_multiphase} (main text)
% isolates why H2Q-BR delivers the headline numbers.  Hidden backlog
% duration drops $18.86\%$, late-hidden prefill queueing drops $12.12\%$,
% and the heavy-tail fragment release span contracts $47.46\%$.  Two
% mechanisms are at work: (i) $Q_S$/$Q_L$ separation prevents a heavy
% hidden continuation from blocking a small answer prefill behind it, and
% (ii) the bounded one-shot release lets an already-spilled long prefill
% make locality-safe progress without reclaiming permanent priority.
% Together they shorten the answer-prefill queue wait that dominates
% aTTFT, while keeping decode service stable enough to bound the aTPOT
% regression at $0.54\%$.

\subsection{SGLang Scheduler vs. vLLM v1 Scheduler}
\label{app:sglang-vllm-scheduler}

\noindent\textbf{Policy contrast.}
The two schedulers differ in what they treat as the first-class unit of
service. SGLang is prefill-first: when a prompt-prefill batch can be
formed, it prefers to build that batch before falling back to decode.
vLLM v1 is running/decode-first: it advances running decode work first
and then admits waiting requests if budget remains. For example, if a
latency-sensitive answer prefill arrives while several decode continuations
are already running, SGLang tries to serve the prefill side first, whereas
vLLM v1 keeps the running decode stream in front of the queue.

\noindent\textbf{Implementation mapping.}
% \begin{sloppypar}
This contrast appears in \path{sota-infer-engine}. In SGLang,
\path{sglang/srt/managers/scheduler.py} attempts prefill before decode
fallback. In vLLM v1, \path{vllm/v1/core/sched/scheduler.py} schedules
running requests before waiting requests. In \sys, the mirror lives
under \path{frontier/scheduler/replica_scheduler}:
\path{sglang_style_replica_scheduler.py} is the prefill-first wrapper,
and \path{vllm_v1_engine_replica_scheduler.py} is the running-first
baseline.
% \end{sloppypar}

\begin{figure}[tbp]
  \centering
  \includegraphics[width=\columnwidth]{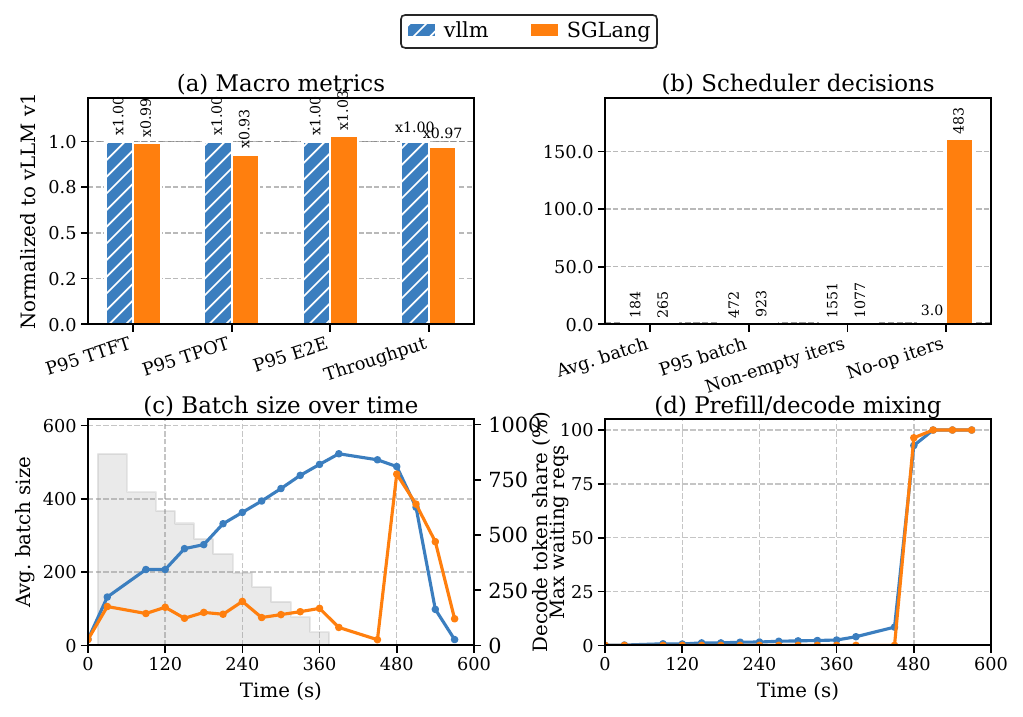}
  \caption{\sys online SharedGPT qps64 scheduler comparison. Panel~(a) reports normalized macro metrics, and the bar labels show multipliers relative to vLLM. Panels~(b)--(d) retain the micro-scheduling view of batch size, backlog, and prefill/decode mixing over time (s).}
  \label{fig:app-sglang-vllm-scheduler}
\end{figure}

\noindent\textbf{Runtime setup.}
To isolate scheduler-policy effects from hardware and model differences, we run \sys in online co-location mode on a simulated 256-H800 deployment of Qwen3-235B-A22B. Both schedulers use the same vLLM-style parallel semantics (\texttt{PP=2}, \texttt{attn\_tp=8}, \texttt{attn\_dp=16}, \texttt{moe\_ep=128}, \texttt{cluster\_num\_replicas=1}), chunked prefill, decode-only CUDA graph capture, disabled prefix caching, and \sys's monolithic MoE stage aggregation path. The active workload is \texttt{sharedgpt\_trace} replay with 1{,}024 requests at qps64; this setting was selected because it sustains a saturated queueing regime while still allowing both schedulers to complete cleanly.

\noindent Figure~\ref{fig:app-sglang-vllm-scheduler} should be read as a normalized comparison: the labels above the bars in panel~(a) report multipliers relative to vLLM, rather than absolute latency scales. In this setting, SGLang remains within 1.06\% of vLLM on P95 TTFT, lowers P95 TPOT by 7.15\%, but increases P95 E2E by 3.01\% and reduces throughput by 2.90\%. The figure therefore does not support a simple scalar ranking; instead, it exposes a trade-off between token-level service regularity and completion-oriented efficiency.

\noindent\textbf{Interpretation.}
The micro panels explain this trade-off at the scheduler level. SGLang increases the average non-empty batch size from 183.703 to 264.554 and the P95 non-empty batch size from 471.5 to 923.0, while reducing non-empty decisions from 1{,}551 to 1{,}077 and increasing no-op decisions from 3 to 483. The saturation envelope remains similar: both schedulers reach a maximum of 868 waiting requests, and their maximum running-request counts stay close (1{,}017 for vLLM and 1{,}021 for SGLang). These counts indicate that the two policies operate under comparable pressure, but allocate that pressure differently across batch construction and admission pacing.

\noindent The time-series view makes the phase structure more explicit. Between 390--420\,s, SGLang's average batch size is 49 with zero decode share, and between 450--480\,s it falls to 15.5 while remaining decode-free; in the same windows, vLLM sustains average batch sizes of 523 and 506.5, with decode shares of 4.0\% and 8.5\%, respectively. The transition into a decode-dominant regime also occurs later for SGLang: at 480--510\,s it reaches an average batch size of 467.450331 with a 96.3902\% decode share, whereas vLLM already reaches 92.8177\% decode share in the same window and moves into full decode dominance one window earlier. This behavior is consistent with a prefill-first two-phase policy that consolidates prompt service before mixing in decode, while vLLM's running-request-first ordering admits decode into the service stream earlier. Under the present 256-H800 SharedGPT setting, that difference explains why SGLang can improve TTFT and TPOT yet still exhibit a small E2E and throughput penalty: the policy shifts work into larger prompt-oriented batches, but postpones mixed-phase service and makes the later decode completion burstier.

\end{document}